\documentclass[11pt]{article}
\pdfoutput=1
\usepackage{jheppub}
\usepackage{graphicx}
\usepackage{amssymb}
\usepackage{amsmath,amssymb}
\usepackage{slashed}
\usepackage{hyperref}
\usepackage{caption}
\usepackage{xcolor}
\usepackage{dsfont}
\usepackage{verbatim}
\usepackage{subfig}

\newcommand\beq{\begin{equation}}
\newcommand\eeq{\end{equation}}

\title{Generalization of QCD$_3$ Symmetry-Breaking and Flavored Quiver Dualities}

\preprint{\today}

\author[a]{Kyle Aitken,}
\author[a]{ Andrew Baumgartner,}
\author[b]{ Changha Choi,}
\author[a]{ Andreas Karch}
\affiliation[a]{Department of Physics, University of Washington, Seattle, WA, 98195-1560, USA}
\affiliation[b]{Physics and Astronomy Department,
Stony Brook University, Stony Brook, NY 11794, USA}

\abstract{We extend the recently proposed symmetry breaking scenario of QCD$_3$ to the so-called ``master'' $(2+1)$d bosonization duality, which has bosonic and fermionic matter on both ends. Using anomaly arguments, a phase diagram emerges with several novel regions. We then construct 2+1 dimensional dualities for flavored quivers using node-by-node dualization. Such dualities are applicable to theories which live on domain walls in QCD$_4$-like theories with dynamical quarks. We also derive dualities for quivers based on orthogonal and symplectic gauge groups. Lastly, we support the conjectured dualities using holographic constructions, even though several aspects of this holographic construction remain mostly qualitative.
}

\begin{document}
\maketitle

\section{Introduction}

In last few years, a new class of conjectured non-supersymmetric dualities have arisen between $(2+1)$-dimensional Chern-Simons theories with fundamental matter. The precise form of these dualities was first written down by Aharony \cite{Aharony:2015mjs},
\begin{subequations}
\label{eq:aharony}
\begin{align}
\label{eq:3d bos}
SU(N)_{-k+N_f/2} \text{ with } N_f \text{ }\psi & \qquad  \leftrightarrow\qquad U(k)_{N}\text{ with } N_f \text{ }\Phi, \\
SU(N)_{-k} \text{ with } N_s \text{ } \phi  & \qquad   \leftrightarrow \qquad U(k)_{N-\frac{N_s}{2}} \text{ with } N_s \text{ } \Psi, \label{eq:3d bos2}
\end{align}
\end{subequations}
where we use ``$\leftrightarrow$'' to mean the two theories flow to the same IR fixed point. Since for each of these dualities one side contains bosons and the other fermions, they are often referred to as ``3d bosonization'' dualities.

It is not yet known how to rigorously show the aforementioned theories are dual, since for a general choice of parameters they are strongly coupled. Nevertheless, there is growing indication of their equivalence.  Significant evidence can be found in the large $N$ and $k$ limit (with $N/k$ fixed) where observables such as the free energy, correlations functions, and operator spectrum have been shown to match across both sides of the duality \cite{Giombi:2011kc, Aharony:2011jz, Minwalla:2015sca, Aharony:2012nh, Jain:2014nza, Inbasekar:2015tsa, Gur-Ari:2016xff}. In the opposite limit when $N=k=N_f=1$, \eqref{eq:aharony} can be used to derive an infinite web of dualities \cite{Karch:2016sxi,Seiberg:2016gmd} among which is the well-known bosonic particle-vortex \cite{Peskin:1977kp,Dasgupta:1981zz} and its recently discovered fermionic equivalent \cite{Son:2015xqa}. Further checks come from anomaly matching across the dualities \cite{Benini:2017dus}, consistency on manifolds with boundaries \cite{Aitken:2017nfd,Aitken:2018joi}, deformations of supersymmetric parents \cite{Kachru:2016aon, Gur-Ari:2015pca, Jain:2013gza}, explicit derivations of related systems from coupled $1+1$ d wires \cite{Mross:2017gny}, and Euclidean lattice constructions \cite{Jian:2018amu, Chen:2017lkr, Son:2018zja, Chen:2018vmz}.

\begin{figure}
\begin{centering}
\includegraphics[scale=0.6]{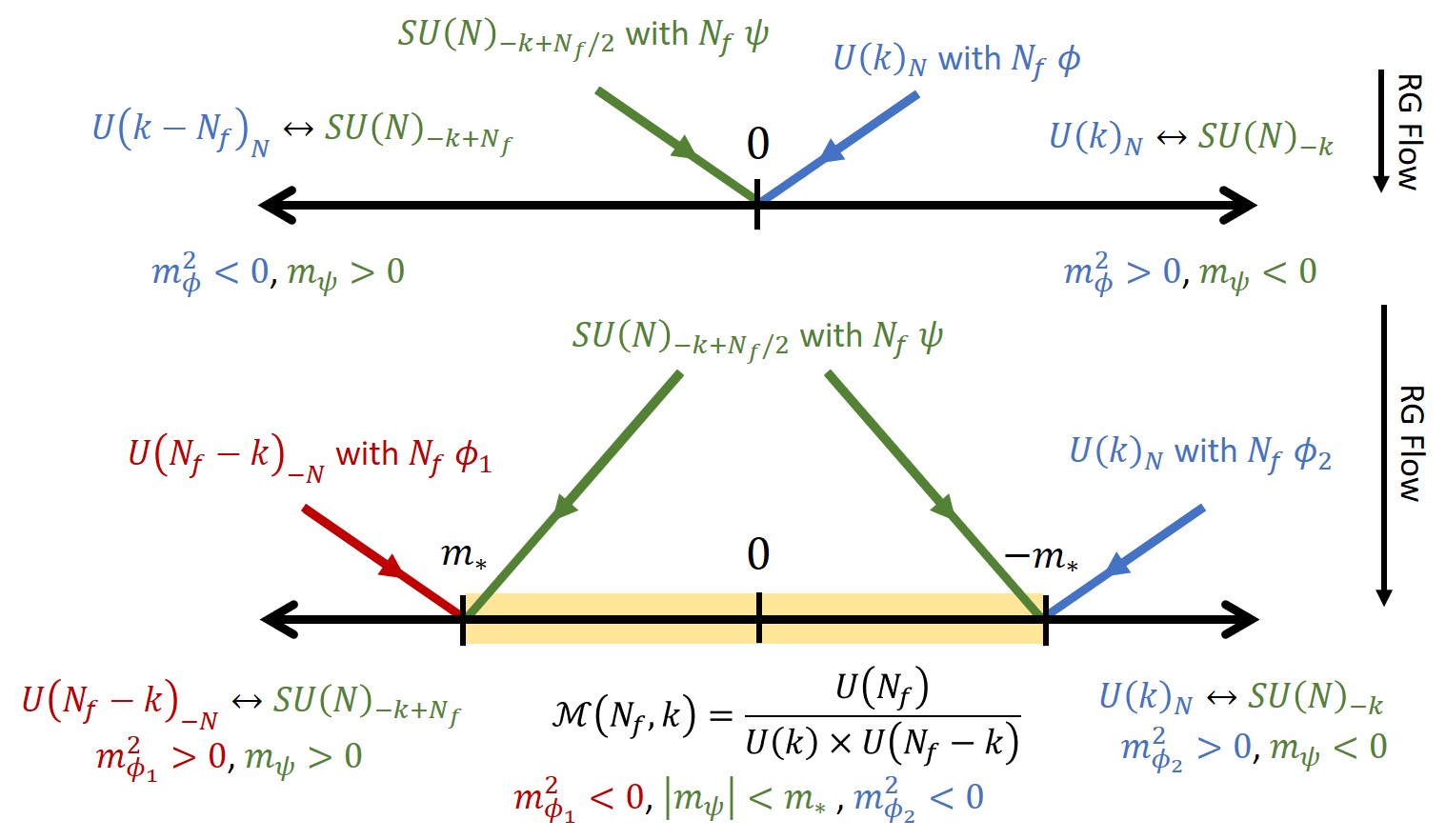}
\par\end{centering}
\caption{\emph{Top:} Mass deformation diagrams of the flavor-bounded 3d bosonization dualities of \eqref{eq:3d bos} where $N_f \leq k$. Here the blue and green lines are the RG flows of the respectively theories, and the black line is the mass deformation axis. \emph{Bottom:} The flavor-violated extension of the usual 3d bosonization dualities proposed by \cite{Komargodski:2017keh}. Here we are working in the regime where $k<N_f<N_*$ and $N>2$. The portion of the axis which is shaded yellow is the Grassmannian phase. The $SU$ levels are \emph{without} the additional $-N_f/2$ shift Ref. \cite{Komargodski:2017keh} (and many others) use to make things look more symmetric. We review the details of this diagram in Sec. \ref{subsec: qcd3 rev}. \label{fig:The-flavor-extension flq}}
\end{figure}

Another consistency check involves mass deforming  each side of the duality. If the mass deformation is large enough, one can integrate out the matter and explicitly show the resulting topological field theories (TFTs) are equivalent via level-rank duality, see Fig. \ref{fig:The-flavor-extension flq}. However, this confirmation breaks down when there are too many flavors of matter. For example, if $N_f > k$ then adding a large negative mass deformation for the scalar on the right hand side of \eqref{eq:3d bos} will completely break the gauge group resulting in a non-linear sigma model. The corresponding mass deformation for the fermionic theory does not appear to exhibit such a phase. Thus, the dualities in \eqref{eq:aharony} were conjectured to only be valid when they satisfy the inequality $N_f\leq k $. We will refer to this as the ``flavor bound".

More recently, a proposal has arisen for an extension of \eqref{eq:3d bos} to the ``flavor-violated'' regime where $k<N_f<N_*(N,k)$ \cite{Komargodski:2017keh} where $N_*$ is some undetermined upper bound. It is conjectured that the fermionic side of \eqref{eq:3d bos} also has a non-linear sigma model phase. For light fermion mass, the fermions form a condensate and spontaneously break their associated flavor symmetry. This means both sides of the duality grow a quantum region described by a complex Grassmannian, see Fig. \ref{fig:The-flavor-extension flq}. We will review the details of this proposal in Sec. \ref{subsec: qcd3 rev}.

Additionally, there has been an extension of the dualities \eqref{eq:aharony} to include fermions and bosons on \emph{both} ends of the duality \cite{Jensen:2017bjo,Benini:2017aed},
\begin{align}
\label{eq:3d master}
SU(N)_{-k+N_f/2} \text{ with } N_f\;\psi \text{ and } N_s\;\phi \qquad\leftrightarrow\qquad U(k)_{N-N_s/2}\text{ with } N_f\;\Phi \text{ and } N_s\;\Psi.
\end{align}
This is the so-called ``master duality'' since  all known flavor-bounded dualities are a special case of said duality. Similar to \eqref{eq:aharony}, this duality is subject to a flavor bound: the mass deformations pass all checks so long as $N_s \leq N$, $N_f \leq k$, with the double saturated case $N_s=N$ and  $N_f=k$ excluded. Note \eqref{eq:3d bos} and \eqref{eq:3d bos2} are simply the $N_s=0$ and $N_f=0$ limits of \eqref{eq:3d master}, respectively.

A natural next step is to combine the two aforementioned extensions of the 3d bosonization dualities. As such, in the first part of this paper we work to extend the master duality to the flavor-violated regime. Naively one may think this extension is fairly trivial and the master duality grows a single new quantum region when fermion mass is light. However, with certain strong coupling assumptions, we find the phase diagrams grows several new quantum regions and exhibits behavior not yet seen in the context of 3d bosonization.

In the second part of this paper we will focus on a particular application of said dualities: flavored quiver dualities applicable to domain walls of QCD$_4$ with $N_F$ fundamental fermions \cite{Gaiotto:2017tne}.\footnote{A quick clarification on notation: throughout this work $N_F$ will be used when referring to the number of flavors of fundamental fermions in $\text{QCD}_4$ while $N_f$ will be used when referring to the parameter in the 3d bosonization dualities.} This is a natural extension of previous work \cite{Aitken:2018cvh}, where some of the present authors proposed a new $2+1$ dimensional duality relating quiver gauge theories to field theory with adjoint matter. Such quiver gauge theories provide a $2+1$ dimensional effective description domain walls and interfaces in $3+1$-dimensional $SU(N)$ Yang-Mills, i.e. the $N_F=0$ case of QCD$_4$.

\begin{figure}[b]
\centering
\includegraphics[scale=0.7]{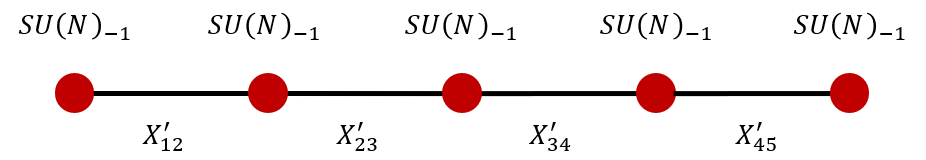}
\caption{Quiver gauge theory for the special case of $n=5$ nodes. As usual, nodes denote gauge groups and links denote bifundamental Wilson-Fisher scalars.} \label{fig:parentcft}
\end{figure}

More specifically the quiver gauge theory
\begin{align}
\label{eq:su_quiver_ym}
[SU(N)_{-1}]^n + \text{bifundamental scalars},
\end{align}
whose precise matter content is displayed in Fig. \ref{fig:parentcft}, was conjectured to be dual to
\begin{align}
\label{eq:u_quiver_ym}
 U(n)_{N} + \text{adjoint scalar}.
\end{align}
Other than simply performing the usual checks, i.e. agreement of massive phases and anomalies, this duality was constructed using two different approaches in \cite{Aitken:2018cvh}: first, following the ideas developed in \cite{Jensen:2017dso}, the original quiver gauge theory was dualized ``node-by-node" using the master duality, \eqref{eq:3d master}. In general, such a procedure will yield a dual quiver gauge theory with the same number of nodes. For the special case where the ranks of the gauge group on all nodes were the same, $N$, the dual gauge theory had $n-1$ confining nodes and so the only surviving gauge group was a single $U(n)_N$ factor, with the bifundamental scalars being replaced with an adjoint ``meson" field. Secondly, the duality was obtained from careful consideration of a holographic embedding of the duality \cite{Jensen:2017xbs}. One starts with the construction of Ref. \cite{Gaiotto:2017tne} that considered interfaces with varying theta angle in four dimensional pure Yang-Mills theory. The TFTs which live on these domain walls depends on the gradient of the theta angle -- we get different TFTs if the gradient is ``shallow" or ``steep".  If the transition between these two regimes is second order, the corresponding fixed point should be governed by the quiver conformal field theory \eqref{eq:su_quiver_ym}. The holographic dual of such interfaces is well known \cite{Witten:1998uka} in the context of the simplest holographic dual for a confining gauge theory, Witten's black hole \cite{Witten:1998zw}. The key point is that the holographic dual at low energies does not give back the quiver gauge theory \eqref{eq:su_quiver_ym}, but its dual incarnation \eqref{eq:u_quiver_ym}.

Both of these techniques can be employed in order to derive more general dualities in the same spirit, even though the holographic construction proves to require a certain amount of additional assumptions. One thing we do in this work is to include fundamental flavors. In \cite{Gaiotto:2017tne} domain walls in $\text{QCD}_{4}$ were considered not just for the case of $N_F=0$ (i.e. pure YM), but also for the cases $N_F=1$ and $N_F>1$, which appear somewhat distinct. The quiver theory described above gets augmented with extra fundamental matter on each node (see Fig. \ref{fig:ferm+su flq}). Once more, we can derive a dual via node-by-node dualization. We will see the $N_F>1$ and $N_F=1$ cases require use of two distinct regimes of the flavor-violated master duality we propose in the first part of this paper.

Holographically, the inclusion of flavor can be accomplished by adding probe $\text{D}8$ and $\overline{\text{D}8}$ branes as in the Sakai-Sugimoto model \cite{Sakai:2004cn}. Holographic theta walls in this context have been discussed recently in \cite{Argurio:2018uup}. While both constructions have their own subtleties, in the end both give closely related conjectured duals for flavored quivers.

A second generalization is to extend the original construction to gauge theories based on orthogonal and symplectic groups. The master duality is known for these gauge groups as well, so once more we can employ a node-by-node dualization. On the holographic side, the projection to orthogonal and symplectic groups can be enforced by orientifolds. Again we see consistency between the node-by-node dualization and the holographic construction.

The organization of this paper is as follows. We first provide a (very) concise review of the master duality in the next subsection, mostly to establish notation used throughout this text. In section \ref{sec:fvm}, we generalize the quantum phase of QCD$_3$ to the case of the \emph{flavor-violated} master duality with careful distinction of the case $N>N_s$ and $N=N_s$, the phase diagrams of which are summarized in figures \ref{fig: master_fv_n>ns} and \ref{fig: master_fv_n=ns}. We also discuss the case of the double-saturated flavor bound. In section \ref{sec:quivers}, we start with the motivation of studying quivers by emphasizing a relation between 3d quiver theories and interfaces transitions in $SU(N)$ QCD$_4$. We analyze and derive the dual quiver theories using node-by-node duality and the holographic construction. We discuss the matching of phases and subtle issue about interaction terms and enhanced flavor symmetries which seems ubiquitous in the flavored-quiver theories. We generalize the previous analysis to the orthogonal and symplectic group in the section \ref{sec:orth and symp}. Finally we summarize our results and comment on the future directions in section \ref{sec:dfd}. Appendices \ref{app:backgrounds} and \ref{sec:Quivers and spinc flq} are devoted to the detailed analysis of background terms and spin$_c$ convention, respectively.

\subsection{Brief Review of Master Duality and Notation}

Let us briefly review the master duality and the notation used throughout this text. Schematically, the master duality conjectures the two sides of \eqref{eq:3d master} are dual in the sense that they flow to the same IR fixed point \cite{Jensen:2017bjo,Benini:2017aed}. At the level of Lagrangians, the master duality is\footnote{This is a modified version of the master duality proposed in \cite{Jensen:2017bjo} with $\tilde{A}_1 \to N\tilde{A}_1$ as well as a flipped BF term and $\tilde{A}_{2}$ coupling on the $U$ side of the duality, but the consistency with a $\text{spin}_{c}$ manifold are unchanged. The BF term which sometimes appears on the $SU$ side of the duality has also been integrated out. In explicit expressions for Lagrangians, we use the convention for the $\eta$-invariant where a positive fermion mass deformation will not change the level. That is, our convention means we take
\begin{align}
i\bar{\psi}\slashed{D}_c \psi -i\left[-\frac{1}{2}\frac{N_f}{4\pi}\text{Tr}_k\left(cdc - i\frac{2}{3}c^3\right)\right] \quad \to \quad i\bar{\psi}\slashed{D}_c \psi.
\end{align}
See \cite{Aitken:2018cvh} for a more thorough review of this duality.}
\begin{subequations}
\begin{align}
\mathcal{L}_{SU} & =\left|D_{b^{\prime}+B+\tilde{A}_{1}+\tilde{A}_{2}}\phi\right|^{2}+i\bar{\psi}\slashed{D}_{b^{\prime}+C+\tilde{A}_{1}}\psi+\mathcal{L}_{\text{int}}-i\left[\frac{N_{f}-k}{4\pi}\text{Tr}_{N}\left(b^{\prime}db^{\prime}-i\frac{2}{3}b^{\prime3}\right)\right]\nonumber \\
 & -i\left[\frac{N}{4\pi}\text{Tr}_{N_{f}}\left(CdC-i\frac{2}{3}C^{3}\right)+\frac{N\left(N_{f}-k\right)}{4\pi}\tilde{A}_{1}d\tilde{A}_{1}+2NN_{f}\text{CS}_{\text{grav}}\right],\\
\mathcal{L}_{U} & =\left|D_{c+C}\Phi\right|^{2}+i\bar{\Psi}\slashed{D}_{c+B+\tilde{A}_{2}}\Psi+\mathcal{L}_{\text{int}}^{\prime} \nonumber \\
& -i\left[\frac{N}{4\pi}\text{Tr}_{k}\left(cdc-i\frac{2}{3}c^{3}\right)-\frac{N}{2\pi}\text{Tr}_{k}(c)d\tilde{A}_{1}+2Nk\text{CS}_{\text{grav}}\right]
\end{align}
\end{subequations}
where the dynamical and background gauge fields with their associated symmetries are given in Table \ref{tab:fields flq}, and the parameters obey $(N_f,N_s)\leq (k,N)$ with the $(N_f,N_s) = (k,N)$ case included. All gauge fields are ordinary connections with the exception of $\tilde{A}_{1\mu}$ and $\tilde{A}_{2\mu}$, which are spin$_c$ connections.
\begin{table}
\begin{centering}
\begin{tabular}{|c|c|c||c|c|c|c|}
\cline{2-7}
\multicolumn{1}{c|}{} & \multicolumn{2}{c||}{\textbf{Gauge Fields}} & \multicolumn{4}{c|}{\textbf{Background Fields}}\tabularnewline
\hline
\textbf{Symmetry} & $SU(N)$ & $U(k)$ & $SU(N_s)$ & $SU(N_f)$ & $U(1)_{m,b}$ ($\text{spin}_{c}$) & $U(1)_{F,S}$ ($\text{spin}_{c}$)\tabularnewline
\hline
\textbf{Field} & $b_{\mu}^{\prime}$ & $c_{\mu}$ & $B_{\mu}$ & $C_{\mu}$ & $\tilde{A}_{1\mu}$ & $\tilde{A}_{2\mu}$\tabularnewline
\hline
\end{tabular}
\par\end{centering}
\caption{Definitions of various gauge fields used in the master duality.\label{tab:fields flq}}
\end{table}

\begin{figure}
\begin{centering}
\includegraphics[scale=0.45]{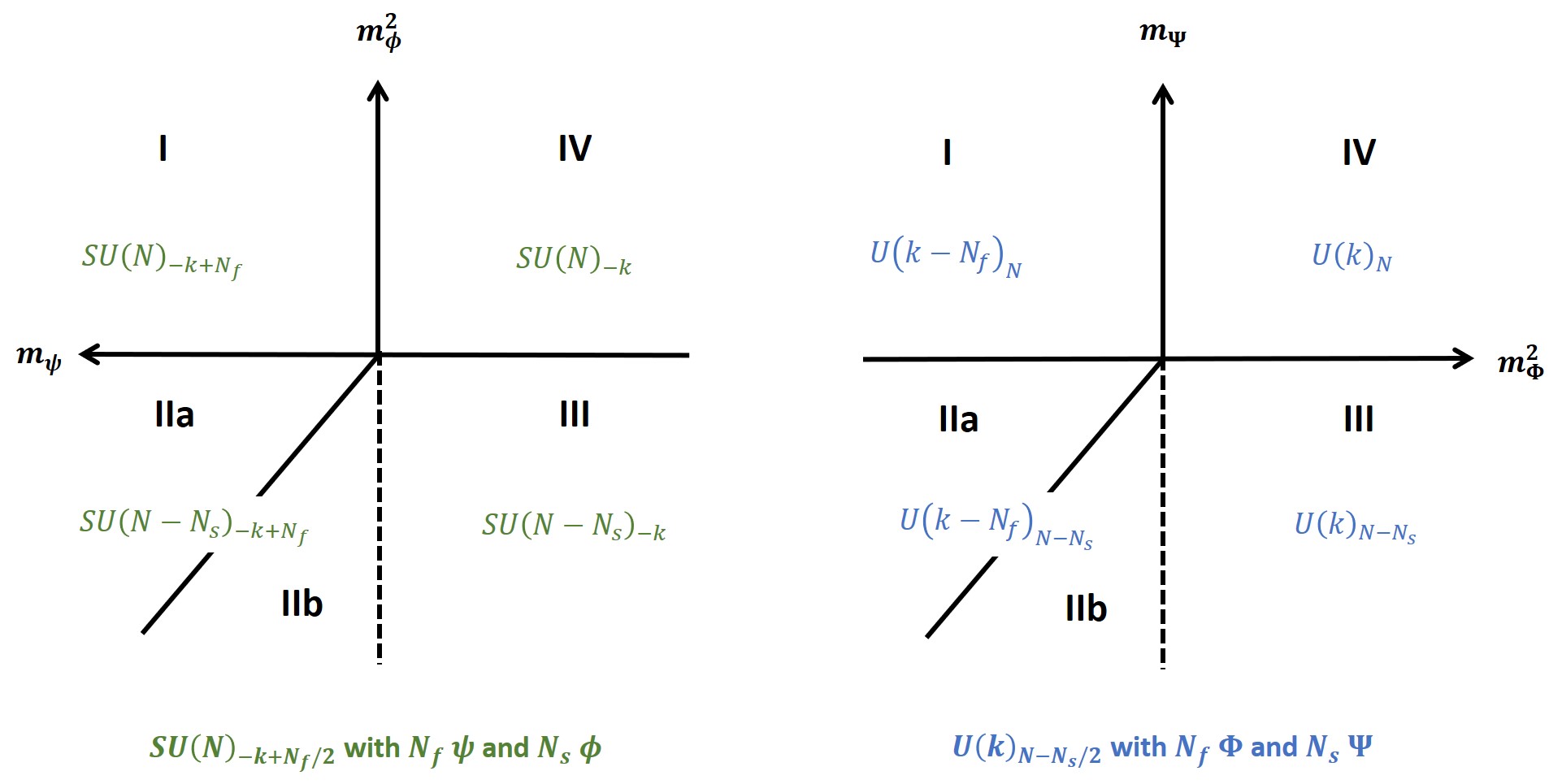}
\par\end{centering}
\caption{Phase diagram of the flavor-bounded master duality. Critical theories separating phases are the black lines. The dotted black lines represent a critical theory that would be present for $N>N_s$, but is not present for $N=N_s$. \label{fig:master-flavor-bounded}}
\end{figure}

The phase diagram for the master duality as a function of scalar and fermion mass deformations is shown in Fig. \ref{fig:master-flavor-bounded}. The $N_s < N$ and $N_s = N$ cases yield topologically distinct phase diagrams. A crucial role is played by the quartic interaction terms, denoted by $\mathcal{L}_\text{int}$ and $\mathcal{L}_\text{int}^\prime$. The scalar$^2$ fermion$^2$ terms in the potential split phase II into two separate phases. This is a result of the fact that the interaction serves as a mass term for the fermions when the scalars acquire a vev. For example, on the $SU$ side the interaction gives rise to an extra phase IIb in which the mass term for some fermions induced by the scalar expectation value via the quartic potential is larger in size but opposite in magnitude when compared to the explicit fermion mass. Without such interactions we would not find a matching of the dualities in phase II.  See \cite{Jensen:2017bjo, Benini:2017aed} for more details. It is useful to think of the interactions as acting unidirectionally -- the scalars can influence the mass of the fermions but not vice versa.

Critical theories lie on the lines separating each of the phases, including the line between IIa and IIb. Such critical theories are constrained by the two phases which they separate via anomaly arguments. The (IIa-IIb) critical theory corresponds to locations where the mass from the interactions exactly cancels that due to the mass deformations, and thus contains massless fermions. These massless fermions which arise from this cancellation mechanism will play a special role in our story, so we call them ``singlet fermions'' and denote them with a subscript $s$, e.g. $\psi_s$.

Note also we can adjust the strength of the interaction terms\footnote{Strictly speaking, the coefficients of the scalar$^2$ fermion$^2$ couplings flow to their IR fixed values and do not correspond to adjustable parameters. When we say we adjust the strength of the coupling we mean that by rescaling the scalars relative to the fermions we zoom into a region of the phase diagram in which the couplings appear to dominate or be unimportant.} which has the effect of moving the location of the diagonal critical line of Fig. \ref{fig:master-flavor-bounded}.  In \cite{Aitken:2018cvh}, we worked in the limit where $SU$ quiver has no interaction terms. This necessitated very large interactions in some of the other theories down the duality chain. For the purposes of the first part of this paper, we will always assume we are working with finite interaction strengths.

\section{Extending the Master Duality}
\label{sec:fvm}

In this section we will argue that the master duality can be extended beyond the flavor bound, similar to the case of QCD$_3$ discussed in \cite{Komargodski:2017keh}. The natural extension is to attempt to look in the regime where $k<N_{f}<N_{*}(N,k)$ and $N_s<N$.

We will begin by reviewing the construction of \cite{Komargodski:2017keh}, which corresponds to the $N_s = 0$ limit. Like the ordinary master duality, the phase diagrams for the $N_s<N$ and $N_s=N$ cases look fairly different due to particular cancellations which occur in the latter case. We will begin with the phase diagram for the more general $N_s<N$ since it is slightly simpler to analyze. The $N=N_s$ case will be considered second and it will be applicable to the quivers we are constructing to model the domain wall behavior of $\text{QCD}_{4}$. Finally, we will discuss extending the master duality to the double saturated case in Sec. \ref{subsec:double-sat}.

\subsection{Review of QCD$_3$ Symmetry-Breaking}
\label{subsec: qcd3 rev}

As was mentioned in the the introduction, the flavor bounds of \eqref{eq:3d bos} are imposed because when $N_f > k$, the Higgs phase of the $U$ side of the theory becomes a non-linear sigma model. This does not appear to be matched on the $SU$ side, because the corresponding fermion mass deformation simply shifts the Chern-Simons level. Hence the bound $N_f\leq k$ was imposed to avoid the mismatch from the non-linear sigma model phase.

The proposal of \cite{Komargodski:2017keh} is that when $k<N_{f}<N_*(N,k)$, for small fermion masses, i.e. $\left|m_{\psi}\right|<m_*$, the $SU$ Chern-Simons theory causes the fermions to condense, yielding a non-linear sigma model whose target space is the complex Grassmannian manifold\footnote{Note, here we are using a slightly different convention than \cite{Komargodski:2017keh} for the levels of the Chern-Simons terms. The $SU$ levels we use do not have the additional $-N_f/2$ shift used in \cite{Komargodski:2017keh} to make the phases look more symmetric.}
\begin{equation}
\mathcal{M}(N_f,k)=\frac{U(N_f)}{U(k)\times U(N_f-k)}.
\label{eq: grassmannian M}
\end{equation}
Said another way, the $SU$ side exhibits ``chiral'' symmetry breaking, yielding as its low energy ``pion'' Lagrangian the target space of \eqref{eq: grassmannian M}. Importantly, it is the number of fermions with $\left|m_{\psi}\right|<m_{*}$ which determines whether or not the associated gauge group confines, resulting in the Grassmannian \eqref{eq: grassmannian M}. Specifically, the assumption is that if the number of light fermions exceeds the bare level of the gauge group (the level without the $\eta$-invariant offset), the fermions still condense. When we construct quivers below we will further split the $N_{f}$ symmetry into smaller subgroups and individually tune their masses. This can lead to a rich structure with many different Grassmannians, but in this work we will generally be concerned with phases outside of this region. More details on such subtleties of the phase diagram can be found in Refs. \cite{Argurio:2019tvw, Baumgartner:2019frr}.

A potential issue with this assumption is that the Grassmannian of the $SU$ side only exists for small mass deformations, see Fig. \ref{fig:The-flavor-extension flq}. This is not matched on the $U$ side if we use the phase diagram of the flavor-bounded duality \eqref{eq:3d bos}, which has a Grassmannian phase for unbounded negative scalar mass deformations. To rectify this, the authors of ref. \cite{Komargodski:2017keh} propose that there are \emph{two} scalar theories dual to the $SU$ end that are patched together to describe the full phase diagram. More explicitly, the duality proposed is
\begin{equation}
SU(N)_{-k+N_{f}/2}\,\text{with }N_{f}\,\psi\qquad\leftrightarrow\qquad\begin{cases}
U(k)_{N}\,\text{with }N_{f}\,\Phi_1 & m_{\psi}=-m_*\qquad\text{(blue)}\\
U(N_f-k)_{-N}\,\text{with }N_{f}\,\Phi_2 & m_{\psi}=m_*\qquad\text{(red)}
\end{cases},\label{eq:flq aharony}
\end{equation}
where the colors in parenthesis correspond to those in Fig. \ref{fig:The-flavor-extension flq} and figures which follow. Note that in both scalars theories, when the mass becomes negative we obtain the same aforementioned Grassmannian manifold of \eqref{eq: grassmannian M}. Accompanying the Grassmannian is a Wess-Zumino-Witten term with a coefficient that is determined by the level of the Chern-Simons theories on the $U$ side. This is important for matching of the operators across the duality, but will not play an essential role in our story. For more details on this term see \cite{Komargodski:2017keh}.
This model is summarized in Fig. \ref{fig:The-flavor-extension flq}.

One feature of this construction that we will see occurs also in the master duality case is that the $SU$ side of the duality does not yield a good description of the quantum phase or the theories which exist at the phase transition. That is, the true IR description on the $SU$ side is hidden by strong dynamics and so we must pass to the $U$ side to obtain a full description of the phase diagram.

\subsection{Flavor-Violated Master Duality}
\label{sec:fvmd}

We now generalize the analysis of \cite{Komargodski:2017keh} to the master duality. In particular, we are interested in mapping out the phase diagram of
\begin{align}\label{eq:master}
SU(N)_{-k+N_f/2} \text{ with } N_f\;\psi \text{ and } N_s\;\phi
\end{align}
when $N_f > k$ and $N_s \leq N$ as a function of the scalar and fermion masses. We will begin our analysis in the asymptotic regimes where the mass deformations are large compared to the strong scale and we don't expect to see quantum phases. When one or both mass deformations are small, we again expect to find phases that may be described by Grassmannian manifolds.

When we give an asymptotically large positive mass to the scalars, we can integrate them out and they have no effect on the gauge group. The theory we are left with is
\begin{equation}
\label{large_scalar_mass}
SU(N)_{-k+N_f/2} \text{ with } N_f \text{ } \psi.
\end{equation}
This is nothing more than the left-hand side of \eqref{eq:flq aharony} and so this portion of the phase diagram is identical to what was found for QCD$_3$. We can do the same for a large negative mass deformation. Assuming the Higgsing is maximal,the gauge group breaks as $SU(N)\to SU(N-N_s)$. The resulting theory is
\begin{equation}
\label{large_neg_scalar_mass}
SU(N-N_s)_{-k+\frac{N_f}{2}} \text{ with } N_f \text{ } \psi
\end{equation}
which again appears to be described by \eqref{eq:flq aharony} so long as $N-N_s>1$.  For now, we are ignoring effects due to the interaction term, but we will return to such considerations when we analyze the Grassmannian regime. We will see such terms make the Grassmannian portion of \eqref{large_neg_scalar_mass} different than that of \eqref{eq:flq aharony} with $N\to N-N_s$.


Since we can match onto \eqref{eq:flq aharony} for the asymptotically large mass phases of the $SU$ side of the master duality, we follow the same reasoning for the $U$ side. We thus assume once more we cannot use a single dual theory to describe all the phases of the $U$ side. This motivates the ``flavor-violated master duality" shown in Fig. \ref{fig: master_fv_n>ns}:
\begin{align}\label{flavor_violated_mast}
\begin{split}
 SU(N)_{-k+N_f/2} & \text{ with }  N_f\;\psi \text{ and } N_s\;\phi \\
 \leftrightarrow & \quad \begin{cases}
U(k)_{N-N_s/2}\text{ with }  N_f\;\Phi_1\text{ and }N_s\;\Psi_1 \quad &m_{\psi}=-m_*\text{  (blue)} \\
U(N_f-k)_{-N+N_s/2}\text{ with }N_f\;\Phi_2\text{ and }N_s\;\Psi_2 \quad & m_{\psi}=m_*\text{  (red)}.
\end{cases}
\end{split}
\end{align}
As a first consistency check, we make sure all four asymptotic phases match across the duality
\begin{subequations}
\begin{align}
\text{(I)}: & & SU(N)_{-k+N_{f}} \qquad & \leftrightarrow\qquad U(N_f-k)_{-N}, \\
\text{(II)}:& & SU(N-N_s)_{-k+N_{f}} \qquad & \leftrightarrow\qquad U(N_f-k)_{-N+N_s},\\
\text{(III)}: & & SU(N-N_s)_{-k} \qquad & \leftrightarrow\qquad U(k)_{N-N_s}, \label{eq:fvmd 3}\\
\text{(IV)}: & & SU(N)_{-k} \qquad & \leftrightarrow\qquad U(k)_N.\label{eq:fvmd 4}
\end{align}
\end{subequations}
The levels for both the dynamical and background fields can be fixed by making sure the mass deformations of the two $U$ theories match that of the $SU$ theories for the asymptotically large deformations we have thus far discussed. The details of this analysis as well as explicit Lagrangians are given in Appendix \ref{app:backgrounds}. Phases I and II follow from the first dual description in  \eqref{flavor_violated_mast}, while phases III and IV correspond to second dual description in \eqref{flavor_violated_mast}. Note that for $N_f\leq k$, the flavor-bounded master duality instead yields the phases
\begin{subequations}
\begin{align}
\text{(I')}: & & SU(N)_{-k+N_f} \qquad & \leftrightarrow\qquad U(k-N_f)_N, \\
\text{(II')}: & & SU(N-N_s)_{-k+N_f} \qquad & \leftrightarrow\qquad U(k-N_{f})_{N-N_s},
\end{align}
\end{subequations}
but phases III and IV are the same as the $N_f > k$ case. We will use this fact to save us some work later on.

The mass mappings between the two sides are slightly complicated by the two scalar descriptions of the $U$ side. For $m_\psi<0$ we have
\begin{align}\label{eq:neg_mass_map}
m_{\psi} + m_* \leftrightarrow -m_{\Phi_1}^2  \qquad m_{\phi}^{2} \leftrightarrow m_{\Psi_1}
\end{align}
while for $m_\psi>0$
\begin{align}\label{eq:pos_mass_map}
m_{\psi} - m_* \leftrightarrow m_{\Phi_2}^2  \qquad m_{\phi}^{2} \leftrightarrow -m_{\Psi_2}.
\end{align}
Once more we see the scalar mass vanishes when the magnitude of the fermion mass is equal to $m_*$.

Now let us discuss what we expect to occur as we move to smaller mass deformations. Once more, we can use \eqref{eq:flq aharony} as a guiding principle for particular phases. Since phases I and IV precisely correspond to the large mass deformations of \eqref{eq:flq aharony}, we expect to find the very same Grassmannian in between them, which we have called phase V. This is indeed consistent with the two $U$ theories of \eqref{flavor_violated_mast}. The story gets slightly more complicated for other regions of the phase diagram. Let us now analyze the $N_s<N$ and $N_s=N$ cases in turn.

\subsubsection*{Flavor-Violation for $N>N_s$}

\begin{figure}
\begin{centering}
\includegraphics[scale=0.45]{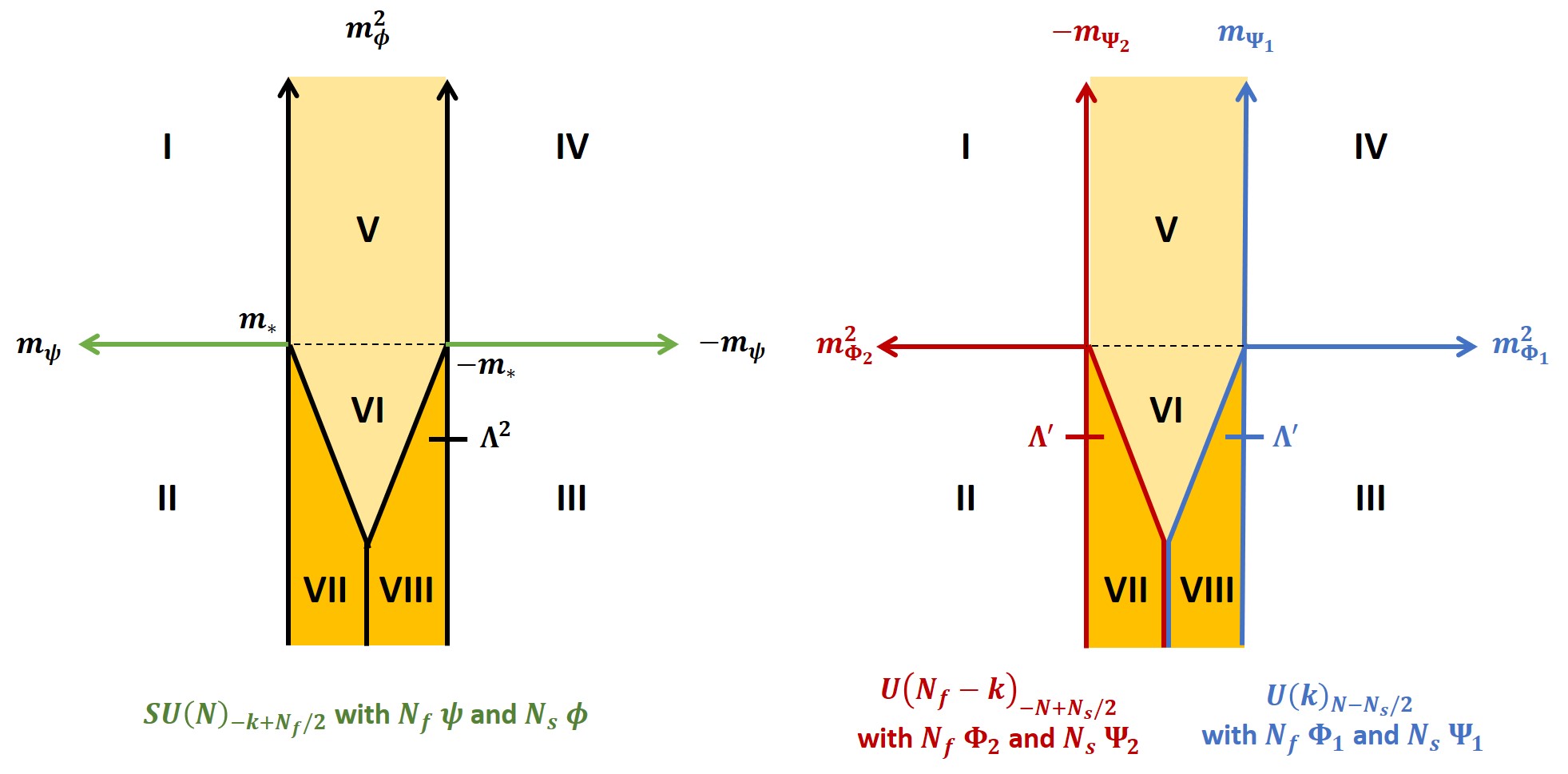}
\par\end{centering}
\caption{$SU$ and $U$ sides of the flavor-violated master duality for $N>N_{s}$. On the $SU$ side, the critical lines in green are well-described by the corresponding $SU$ theory. The critical lines in black are best described by the $U$ duals. \label{fig: master_fv_n>ns}}
\end{figure}

Summarizing our results first, the phase diagram of \eqref{eq:master} and the claimed duals for $N>N_{s}$ are given in Fig. \ref{fig: master_fv_n>ns}. As we have learned, many aspects of the phase diagram are better described on the $U$ side, so much of what is drawn on the $SU$ is learned by considering how the two dual $U$ theories can be consistent in the presence of a finite interaction term. Phases I, V, and IV correspond to the three phases of \eqref{eq:flq aharony} conjectured by Ref. \cite{Komargodski:2017keh}.

The $SU$ phase diagram contains the following phases
\begin{subequations}
\begin{align}
\text{(I)}:\qquad & SU(N)_{-k+N_{f}} \times \left[SU(N_f)_N\times SU(N_s)_0  \times J_\text{I} \right] \\
\text{(II)}:\qquad & SU(N-N_s)_{-k+N_{f}} \times \left[SU(N_f)_N\times SU(N_s)_{-k+N_f} \times J_{\text{II}}\right] \\
\text{(III)}:\qquad & SU(N-N_s)_{-k} \times \left[SU(N_f)_0\times SU(N_s)_{-k} \times J_{\text{III}}\right]\\
\text{(IV)}:\qquad & SU(N)_{-k} \times \left[SU(N_f)_0\times SU(N_s)_0 \times J_{\text{IV}}\right] \\
\text{(V)\text{ to }(VIII)}:\qquad & \text{Better described by $U$ side}
\end{align}
\label{eq:su phase n>ns}
\end{subequations}
where the Chern-Simons theories in $[\cdots]$ belong to background gauge fields. Meanwhile, the $U$ side is given by
\begin{subequations}
\begin{align}
\text{(I)}:\qquad & U(N_f-k)_{-N} \times \left[SU(N_f)_N\times SU(N_s)_0 \times J_\text{I}\right] \\
\text{(II)}:\qquad & U(N_f-k)_{-N+N_s} \times \left[SU(N_f)_N\times SU(N_s)_{-k+N_f} \times J_{\text{II}}\right] \\
\text{(III)}:\qquad & U(k)_{N-N_s} \times \left[SU(N_f)_0\times SU(N_s)_{-k} \times J_{\text{III}}\right]\\
\text{(IV)}:\qquad & U(k)_N \times \left[SU(N_f)_0\times SU(N_s)_0 \times J_{\text{IV}} \right] \\
\text{(V),\;(VI)}:\qquad & \mathcal{M}(N_f,k)\times \left[SU(k)_N\times SU(N_f-k)_0\times SU(N_s)_0 \times J_{\text{V,VI}}\right]\\
\text{(VII)}:\qquad & \mathcal{M}(N_f,k)\times \left[SU(k)_N\times SU(N_f-k)_{N_s}\times SU(N_s)_{N_f-k} \times J_{\text{VII}}\right]\\
\text{(VIII)}:\qquad & \mathcal{M}(N_f,k)\times \left[SU(k)_{N-N_s}\times SU(N_f-k)_0\times SU(N_s)_{-k} \times J_{\text{VIII}}\right].
\end{align}
\label{eq:u phase n>ns}
\end{subequations}
We label the critical theories by the two phases which they separate. They are
\begin{subequations}
\begin{align}
\text{(I-II)}:\qquad & U(N_f-k)_{-N+N_s/2} \text{ with } N_s \text{ }\Psi_2 \quad \leftrightarrow \quad SU(N)_{-k+N_{f}} \text{ with } N_s \text{ }\phi \\
\text{(III-IV)}:\qquad &U(k)_{N-N_s/2}  \text{ with } N_s \text{ }\Psi_1 \qquad \qquad  \leftrightarrow \qquad SU(N)_{-k}  \text{ with } N_s \text{ }\phi\\
\text{(I-V)} : \qquad & U(N_f-k)_{-N} \text{ with } N_f \text{ }\Phi_2 \\
\text{(IV-V)} : \qquad & U(k)_{N} \text{ with } N_f \text{ }\Phi_1 \\
\text{(II-VII)}:\qquad & U(N_f-k)_{-N+N_s} \text{ with } N_f \text{ }\Phi_2 \\
\text{(III-VIII)}:\qquad & U(k)_{N-N_s}  \text{ with } N_f \text{ }\Phi_1 \\
\text{(VI-VII)}:\qquad & \mathcal{M}(N_f,k) \text{ with } (N_f-k)N_s \text{ }\psi_s\\
\text{(VI-VIII)}:\qquad & \mathcal{M}(N_f,k)  \text{ with } kN_s \text{ }\psi_s^\prime\\
\text{(VII-VIII)}:\qquad & \mathcal{M}(N_f,k)  \text{ with } (N_f-k)N_s \text{ }\psi_s \text{ and } kN_s \text{ }\psi_s^\prime.
\end{align}
\end{subequations}
Most of the critical theories are described by $U$ side alone except (I-II) and (III-IV) which have an $SU$ description related by \eqref{eq:aharony}. As a reminder, $\psi_s$ and $\psi_s^\prime$ denote neutral fermions in the phases where dynamical gauge group is completely broken. By construction, each of the critical theories correctly describes the transition between the two phases which it separates, consistent with anomaly constraints.

The $J_i$ for $i= \text{I, II, III, IV}$ are the Abelian Chern-Simons levels of the two $U(1)$ background gauge fields $\tilde{A}_1$ and $\tilde {A}_2$ in the $i$-th phase. When one enters the Grassmannian phases, the breaking of the $SU(N_f)$ global symmetry yields an additional background $U(1)$ field, which we call $\tilde{A}_3$. We write these $U(1)$ levels as
\begin{align}
\label{ eq: U(1) definition original}
J_i \equiv J_i^{ab} \frac{1}{4\pi} \tilde A_a d \tilde A_b
\end{align}
where the $J^{ab}_i$ are\footnote{We suppress the third column/row for phases outside of the Grassmannian regime for brevity. For more details on how these terms are calculated, see \cite{Jensen:2017bjo,Benini:2017aed,Aitken:2018joi}.}


\begin{subequations}
\begin{align}
 & J_{\text{I}}^{ab}=\begin{pmatrix} N(N_f-k) & 0 \\ 0 & 0 \end{pmatrix}
\\ & J_{\text{II}}^{ab}=\frac{N(N_f-k)}{N-N_s }\begin{pmatrix} N & N_s\\ N_s & N_s \end{pmatrix}
\\ & J_{\text{III}}^{ab}=\frac{-Nk}{N-N_s }\begin{pmatrix} N & N_s\\ N_s & N_s \end{pmatrix}
\\ & J_{\text{IV}}^{ab}= \begin{pmatrix} -Nk  & 0 \\ 0 & 0 \end{pmatrix}
\\ & J_{\text{V,VI}}^{ab}= \begin{pmatrix} 0 & 0 & -Nk \\ 0 & 0 & 0 \\ -Nk & 0 & Nk\end{pmatrix}
\\& J_{\text{VII}}^{ab} = \begin{pmatrix}0 & 0 & -Nk \\ 0 & N_s(N_f-k) & -N_s k \\ -Nk & -N_s k & \frac{N_s k^2}{N_f-k}+Nk \end{pmatrix}
\\& J_{\text{VIII}}^{ab} = \begin{pmatrix} 0 & 0 & -Nk \\ 0 & -N_s k & -N_s k \\ -Nk & -N_s k & (N-N_s)k\end{pmatrix}.
\end{align}
\label{eq:abelian factors original basis}
\end{subequations}
See appendix \ref{app:backgrounds} for more detailed discussion about the Lagrangian and background fields. We now explain in more detail how we determined these phases and critical lines. 

\paragraph*{Construction of Flavor-Violated Master Duality}

As mentioned above, we follow the guiding principle of Ref. \cite{Komargodski:2017keh} to conjecture the phase diagram on the $U$ side of the duality. For Aharony's duality, the natural way of constructing the $U$ side of the flavor-violated phase diagram was to overlay the phase diagram of two scalar theories, which was consistent since both scalar theories exhibited the same non-linear sigma model phase. It follows that as one traverses the quantum phase of the $U$ side one must switch over from one scalar description to another. This is the same principle by which we have constructed the phase diagram for the master duality.

What complicates the phase diagram description for the flavor-violated master duality is the presence of the interaction term, which leads to an additional splitting of phases. It is straightforward to see the additional critical line from the interaction term is present in what will become the overlapping Grassmannian phases of the $U$ theories. Since phases I, V, and IV should correspond to \eqref{eq:flq aharony}, the new critical line should have no effect deep into said regions. One can also verify that the overlapping scalar theories of \eqref{eq:master} are not consistent outside of phase V without an interaction term. Per these constraints, we conjecture the interaction term's coefficients are chosen such that it splits the Grassmannian region below phase V. Furthermore, since such critical lines cannot simply terminate, at the crossover between the two scalar theories we conjecture the two critical lines merge. Rather remarkably, the merged critical line VII-VIII is precisely the theory needed to be consistent under anomaly constraints given the two phases it separates.

More explicitly, to reach a consistent intermediate phase we require the interaction for the blue theory to have a positive coefficient for its interaction term so as to give the fermions a positive mass when $\Phi_1$ is Higgsed. The interaction takes the form $\mathcal{L}_{\text{int}}^{1}\supset c^{\prime}\left(\Phi_{1}^{\dagger\alpha M}\Psi_{\alpha N}^1\right)\left(\bar{\Psi}_1^{\beta N}\Phi_{\beta M}^1\right)$ with $\alpha,\beta$ indices for the $U(k)$ gauge field, $M$ an index for the $SU(N_f)$ global symmetry, $N$ index for the $SU(N_s)$ global symmetry, and $c^{\prime}>0$. We then must give the fermions a negative mass to make the singlets light. This gives $kN_s$ light fermions living on the critical theory between phases VI and VIII.  Since the axis of the fermion mass deformation of the red theory is flipped relative to the blue one, it requires a negative coefficient for the interaction term, and so $\mathcal{L}_{\text{int}}^{2}\supset-c^{\prime}\left(\Phi_{2}^{\dagger\alpha M}\Psi_{\alpha N}^2\right)\left(\bar{\Psi}_2^{\beta N}\Phi_{\beta M}^2\right)$ with $\alpha,\beta$ now $U\left(N_{f}-k\right)$ indices. This gives $(N_f-k)N_s$ light fermions on the critical line between phases VII and VI. At the crossover between (VI-VII) and (VI-VIII), the two critical theories unite into $N_fN_s$ light fermions, which is consistent with the transition from VII to VIII. This separates the Grassmannian phase into the three different regimes shown in Fig. \ref{fig: master_fv_n>ns}. These phases all share the same Grassmannian $\mathcal{M}(N_{f},k)$, but have distinct non-Abelian background Chern-Simons terms. \footnote{Also note that phase VII and VIII have a different coefficient of the Wess-Zumino-Witten term from phases V and VI.  This comes from the Chern-Simons level of the $U$ gauge theories bordering the Grassmannian phases.}

For the rest of the phase diagram, calculating the critical theories is straightforward. As conjectured in \cite{Komargodski:2017keh}, the critical lines corresponding to I-V and IV-V, as well as II-VII and III-VIII occur at some finite fermion mass deformation $\pm m_{*}$. These are obtained by application of  \eqref{eq:flq aharony}. As such these lines are better described by the $U$ side of the theory. This is different from the I-II and III-IV critical lines, which follow from a straight forward application of \eqref{eq:3d bos2}.

Although much of this phase diagram lives in the strong coupling regime, it offers a plausible scenario for the matching of the dynamical and background gauge terms, which we discuss explicitly in Appendix \ref{app:backgrounds}. The gravitational Chern-Simons terms can also be calculated and are consistent with expectations from level-rank duality.

\subsubsection*{Flavor-Violation for $N_s = N$}

\begin{figure}
\begin{centering}
\includegraphics[scale=0.45]{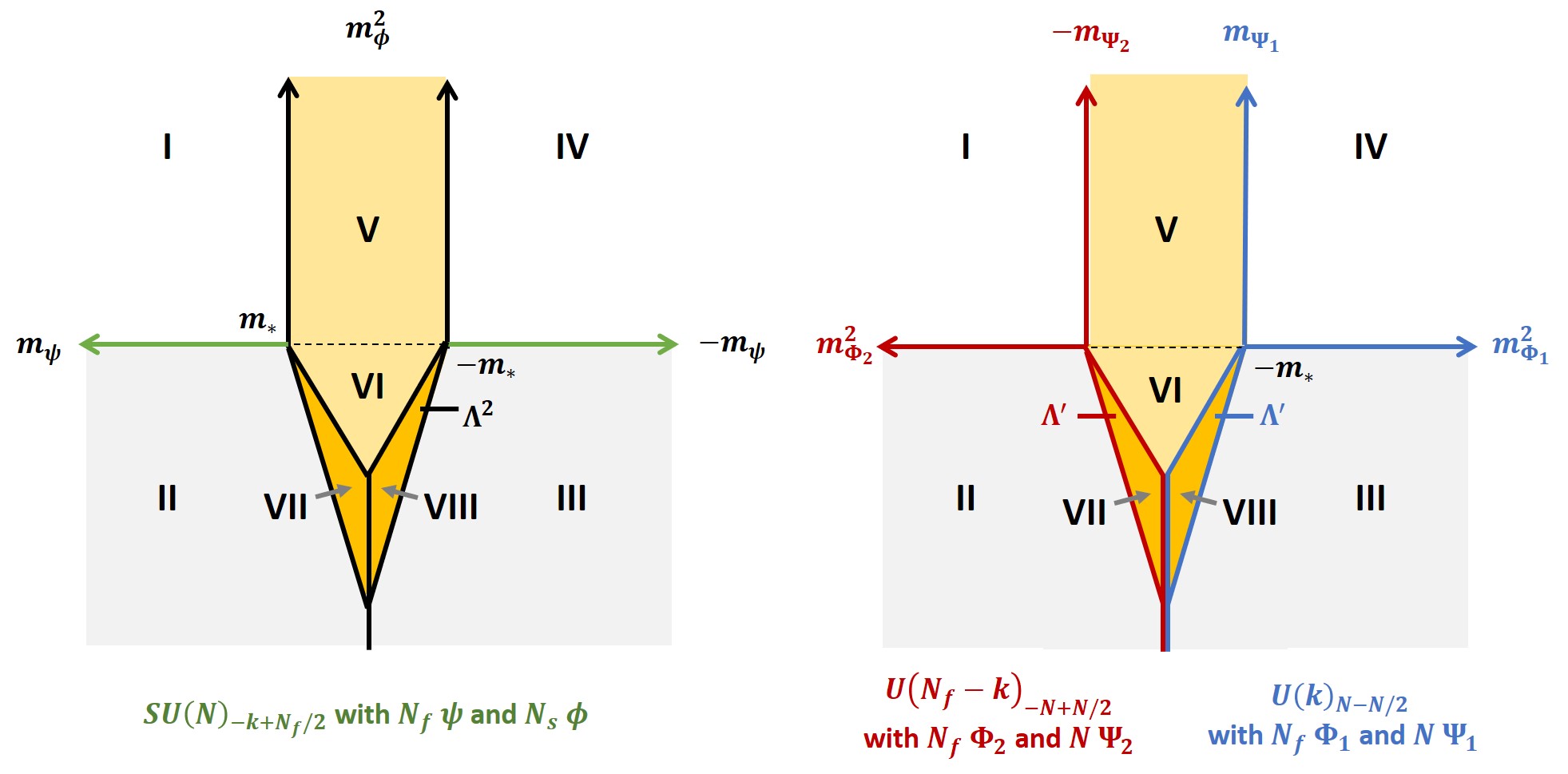}
\par\end{centering}
\caption{$SU$ and $U$ sides of the flavor-violated master duality for $N=N_s$ and $|k|<N_f<N_*$. \label{fig: master_fv_n=ns}}
\end{figure}

The conjectured phase diagram for $N=N_s$ is shown in Fig. \ref{fig: master_fv_n=ns}, with the corresponding phases given in \eqref{eq:su phase n=ns} and \eqref{eq:u phase n=ns}.

To see why $N=N_s$ needs to be considered separately, we begin by looking at the $SU$ side. In the phase where $m_{\phi}^2<0$, the gauge group is $SU(N-N_s)$ and is thus completely broken when $N=N_s$. Hence for large enough scalar mass, we assume the strong dynamics of the gauge groups which was responsible for the condensation of the fermions is eliminated. If this occurs, the Grassmannian phase cannot persist as we approach $m_\phi^2 \to -\infty$. We will take the termination to occur around $m_{\phi}^{2} \sim \Lambda^2$. When one goes beyond this scale, we assume the gauge bosons of $SU(N)$ pick up too large of a mass (relative to the mass from the Chern-Simons terms) to cause the fermions to condense, meaning the Grassmannian phase terminates as one goes $\left|m_{\phi}^{2}\right|\gg\Lambda^2$.  Note that this is very different from the $m_\phi^2>0$ phase, where the scalars are simply gapped out and the fermion condensate is unaffected.


This should be mirrored on the $U$ side of the duality for $N=N_s$, where now we have the special case where the level of the $U$ gauge group is zero and thus the gauge bosons are truly massless in the classical theory. Since we are in the strongly coupled regime we can only make conjectures, but we offer a plausible mechanism to match the $SU$ side.  Like the $SU$ side, we will make the assumption that the termination comes when the matter, this time the fermion, has a large enough mass (in magnitude) to be integrated out. We conjecture that for large enough fermion mass, the level being $N-N_s=0$ means the $U$ gauge group is confining and it induces a mass gap for the scalars. This gap is large, but can be cancelled off by an appropriate mass deformation, similar to the singlet fermion picture described above. This allows us to still have a critical theory of light scalars along the II-VII and VIII-III transition that can drive us into the Grassmannian and give a mass to the singlet fermions.  Once these theories become degenerate, i.e. when the two lines of light scalar meet in Fig. \ref{fig: master_fv_n=ns} the description breaks down and the Grassmannian terminates, leaving just the singlet fermions on the II-III transition.

The phases of the $SU$ side are\footnote{Here we use the fact that both the theories $SU(0)$ and $U(k)_{0}$ lead to a decoupled $U(1)_0$ theory. To see this on the $SU$ side, note the gauge group is completely broken down by the scalar vacuum expectation value, but this leaves a single light degree of freedom which is the Goldstone boson associated with the spontaneous breaking of the $U(1)$ global symmetry. On the $U$ side, the Chern-Simons term disappears and thus we are left with $U(k)$ Yang-Mills in the IR. The $SU(k)$ part of this confines, leaving the Abelian part of the gauge group which has a light photon. The photon is also associated with the spontaneously broken $U(1)$ symmetry corresponding to conserved particle flux (this is easiest to see in the dual photon language). Hence both of these light degrees of freedom can be associated with a spontaneous symmetry breaking of the $U(1)$ global symmetry (i.e. the one associated with $\tilde B=\tilde{A}_1+\tilde A_2$).}
\begin{subequations}
\begin{align}
\text{(I)}:\qquad & SU(N)_{-k+N_{f}} \times \left[SU(N_f)_N\times SU(N_s)_0  \times J_\text{I} \right] \\
\text{(II)}:\qquad & U(1)_{0} \times \left[SU(N_f)_N\times SU(N_s)_{-k+N_f} \times U(1)_{N(N_f-k)} \right] \\
\text{(III)}:\qquad & U(1)_0 \times \left[SU(N_f)_0\times SU(N_s)_{-k} \times U(1)_{-Nk} \right]\\
\text{(IV)}:\qquad & SU(N)_{-k} \times \left[SU(N_f)_0\times SU(N_s)_0 \times J_\text{IV} \right] \\
\text{(V)\text{ to }(VIII)}:\qquad & \text{Better described by $U$ side.}
\end{align}
\label{eq:su phase n=ns}
\end{subequations}
As we saw for Aharony's duality, the $U$ side is a better description for certain phases. Phases of the $U$ side is given by
\begin{subequations}
\begin{align}
\text{(I)}:\qquad & U(N_f-k)_{-N} \times \left[SU(N_f)_N\times SU(N_s)_0 \times  J_\text{I} \right] \\
\text{(II)}:\qquad & U(1)_0 \times \left[SU(N_f)_N\times SU(N_s)_{-k+N_f} \times  U(1)_{N(N_f-k)} \right] \\
\text{(III)}:\qquad & U(1)_0 \times \left[SU(N_f)_0\times SU(N_s)_{-k} \times  U(1)_{-Nk}\right]\\
\text{(IV)}:\qquad & U(k)_N \times \left[SU(N_f)_0\times SU(N_s)_0 \times  J_\text{IV} \right] \\
\text{(V),\;(VI)}:\qquad & \mathcal{M}(N_f,k)\times \left[SU(k)_N\times SU(N_f-k)_0\times SU(N_s)_0 \times  J_\text{V}\right] \\
\text{(VII)}:\qquad & \mathcal{M}(N_f,k)\times \left[SU(k)_N\times SU(N_f-k)_{N_s}\times SU(N_s)_{N_f-k} \times J_\text{VII}\big|_{N=N_s}\right]\\
\text{(VIII)}:\qquad & \mathcal{M}(N_f,k)\times \left[SU(k)_{0}\times SU(N_f-k)_0\times SU(N_s)_{-k} \times  J_\text{VIII}\big|_{N=N_s} \right].
\end{align}
\label{eq:u phase n=ns}
\end{subequations}
We now have a single $U(1)$ background gauge field for phases II and III, which is $\tilde{A} \equiv \tilde A_1 = - \tilde A_2$. This is because the global $U(1)$ baryon/monopole symmetry is spontaneously broken in the phases II and III for $N=N_s$ (see Appendix \ref{app:backgrounds} for more details).

Finally, critical theories are given by
\begin{subequations}
\begin{align}
\text{(I-II)}:\qquad & U(N_f-k)_{-N/2} \text{ with } N \text{ }\Psi_2 \quad\leftrightarrow \quad SU(N)_{-k+N_{f}} \text{ with } N \text{ }\phi \\
\text{(III-IV)}:\qquad &U(k)_{N/2}  \text{ with } N \text{ }\Psi_1 \qquad\qquad \leftrightarrow \quad SU(N)_{-k}  \text{ with } N \text{ }\phi\\
\text{(I-V)} : \qquad & U(N_f-k)_{-N} \text{ with } N_f \text{ }\Phi_2 \\
\text{(IV-V)} : \qquad & U(k)_{N} \text{ with } N_f \text{ }\Phi_1 \\
\text{(II-VII)}:\qquad & U(N_f-k)_{0} \text{ with } N_f \text{ }\Phi_2 \\
\text{(III-VIII)}:\qquad & U(k)_{0}  \text{ with } N_f \text{ }\Phi_1 \\
\text{(VI-VII)}:\qquad & \mathcal{M}(N_f,k) \text{ with } (N_f-k)N \text{ }\psi_s\\
\text{(VI-VIII)}:\qquad & \mathcal{M}(N_f,k)  \text{ with } kN \text{ }\psi_s^\prime\\
\text{(VII-VIII)}:\qquad & \mathcal{M}(N_f,k)  \text{ with } (N_f-k)N \text{ }\psi_s \text{ and } kN \text{ }\psi_s^\prime \\
\text{(II-III)}:\qquad & U(1)_0  \text{ with decoupled } NN_f \text{ }\tilde \psi_s.
\end{align}
\end{subequations}
In the above, $\psi_s, \psi_s^\prime$ and $\tilde \psi_s$ denote neutral fermions in the phases where dynamical gauge group is completely broken. Note that in contrast to $N>N_s$ case, here we have critical line (II-III) which can only be described by $SU$ side using semiclassical analysis.

\subsection{Double-Saturated Flavor Bound}
\label{subsec:double-sat}

We now investigate the validity of the doubly saturated flavor bound, which will be applicable to the flavored quiver we construct in the next section. That is, can the master duality be extended to also hold for the case $\left(N_{s},N_{f}\right)=\left(N,k\right)$?

Let us review in more detail why the master duality was invalid in such a limit. In \cite{Jensen:2017bjo} it is argued that the phases (shown in Fig. \ref{fig:master-flavor-bounded}) for $N_{s}=N$ are
\begin{subequations}
\begin{align}
\text{(I)}: & & SU(N)_0 \qquad & \leftrightarrow \qquad U(0)_{N} \\
\text{(IIa)}: & & U(1)_{0} \qquad & \leftrightarrow\ \qquad U(0)_{0} \\
\text{(IIb)}: & & SU(0)_{0} \qquad &  \leftrightarrow \qquad U(0)_{0} \\
\text{(III)}: & & U(1)_{0} \qquad & \leftrightarrow\ \qquad U(1)_{0} \\
\text{(IV)}: & & SU(N)_{-k} \qquad & \leftrightarrow\ \qquad U(k)_{N}
\end{align}
\end{subequations}
with mass mapping $m_{\phi}^{2}\leftrightarrow m_{\Psi}$ and $m_{\psi}\leftrightarrow-m_{\Phi}^{2}$. Phase I, III, and IV should all match, since these phases reduce to Aharony's dualities, \eqref{eq:aharony}, and we know the flavor saturated cases pass all tests in those dualities. There appears to be a conflict already in phase I, where one side of the theory is a completely broken $U(0)$ theory, while the other end of the theory is simply $SU(N)$ Yang-Mills with no Chern-Simons term. By assumption, the latter of these confines, so the low energy limit is empty. Meanwhile, in the completely broken theory, the lightest excitations are finite mass vortices.  Thus there are no light degrees of freedom in either case, which is consistent. The matching in phase III occurs because each side is described by a decoupled $U(1)_0$, as we saw earlier in the $N=N_s$ flavor violated master.

The conflict in the double saturated master duality then arises in phase IIa and IIb. Let us compare phase IIa to what we had in phase III. On the $SU$ side, we have gone from $SU(0)_{-k}$ to $SU(0)_{0}$. The breaking and subsequent Goldstone boson which occurred in phase III was irrespective of the Chern-Simons level, so there is no difference in the resulting light degrees of freedom in this phase. Meanwhile on the $U$ side we have gone from $U(k)_{0}$ to $U(0)_{0}$. When the rank of the gauge group is reduced to zero, there is no light photon left in the theory. Thus we have a mismatch in the degrees of freedom on either side, leading Refs. \cite{Benini:2017aed,Jensen:2017bjo} to postulate that the master duality no longer holds for such a case.

To summarize the issue: changing the \emph{level} on the $SU$ side has no effect on the theory, but changing the \emph{rank} does affect the light degrees of freedom on the $U$ side of the theory, but only for the case of $N_{f}=k$ where the rank of the $U$ group is reduced to zero. As we mentioned above, the light degrees of freedom in this phase should be associated with a spontaneous symmetry breaking of the $U(1)$ global symmetry associated to $\tilde{A}_{1}$.

To extend the validity of the master duality to the double saturated case is simple -- we choose to \emph{explicitly} break the $U(1)$ global symmetry on both sides of the duality. This will eliminate the Goldstone bosons associated with the spontaneous breaking of said symmetry, and thus we will again have a matching in phases IIa and IIb. Note this will also modify the light degrees of freedom in said phases.

Perhaps not by coincidence, we have explicitly broken the same $U(1)$ global symmetry in the construction of the $SU(N)$ Yang-Mills $n$-node quivers \cite{Aitken:2018cvh}. Specifically, we introduced a $\det\left(Y_{i,i+1}\right)$ terms and appropriate monopole term on the $U$ side to eliminate the $U(1)^{n-1}$ symmetry. It is straightforward to see this will be necessary for our $N_{F}\geq1$ quivers as well, since there we would also like to prevent the additional Goldstone bosons arising. Hence, by eliminating these $U(1)$ global symmetries on both sides of the duality, we have also extended the master duality to validity in the $(N_{f},N_{s})=(k,N)$ case. This allows the master duality to be valid for the $N_{F}=1$ quivers, which we construct in the next section.

\section{Adding Flavors to Quivers}
\label{sec:quivers}

We now turn our attention to the construction and dualization of flavored quiver gauge theories. These purely $2+1$ dimensional theories will serve as an effective description of interfaces in $3+1$ dimensional QCD when one varies the $\theta$ angle along a particular coordinate direction. Interestingly anomaly considerations alone aren't sufficient to pin down the $2+1$ dimensional theory. While it is difficult to prove, it has been argued in \cite{Gaiotto:2017tne} that different theories govern the steep versus the shallow interface.  Concretely, let us focus on the case where the $\theta$ angle experiences a net jump of $2 \pi n$ with an integer $n$. In this case the shallow interface in a 3+1 dimensional gauge theory with gauge group $SU(N)$ is believed to be described by a 2+1 dimensional $[SU(N)_{-1} ]^n$ gauge theory, whereas the steep wall is described by a single $SU(N)_{-n}$ theory.

\begin{figure}
\begin{centering}
\includegraphics[scale=0.5]{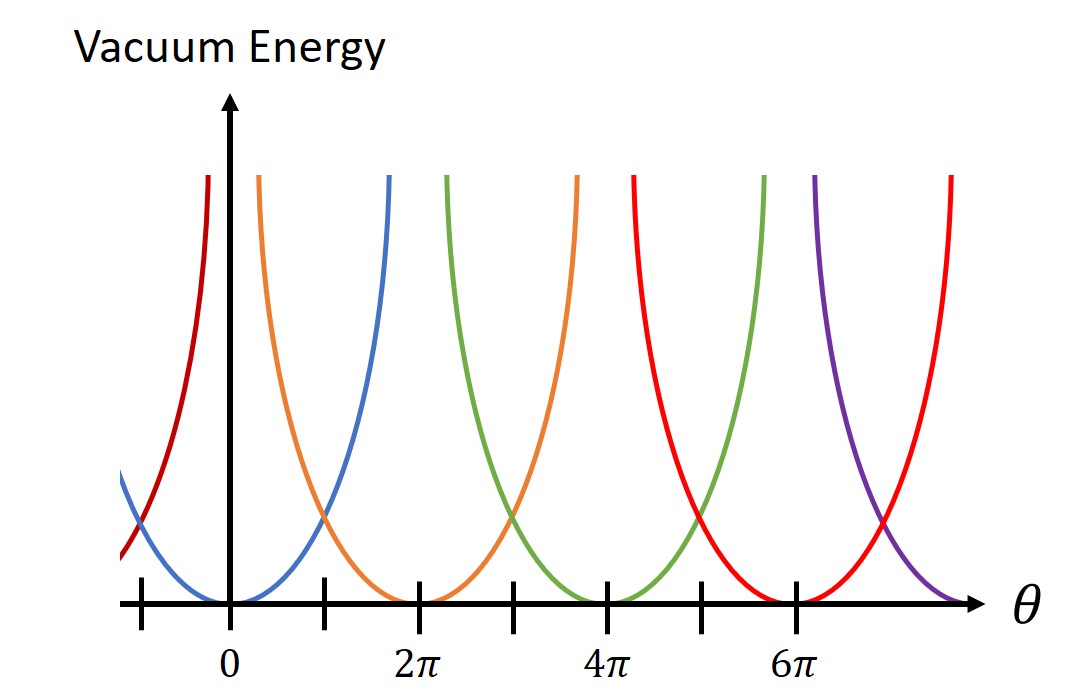}
\par\end{centering}
\caption{Expected behavior of the vacuum energy as a function of $\theta$ in a large $N$ gauge theory. Distinct branches are shown in different colors. \label{fig:thetapotential}}
\end{figure}

At least for the shallow interface this can easily be argued based on the general expectations for the $\theta$ dependence in $SU(N)$ gauge theories, at least at large $N$, as described in \cite{Witten:1998uka}. The vacuum energy in any given vacuum at large $N$ can be shown to be $2 \pi N$ periodic. In order to reconcile this with the expected $2 \pi$ periodicity of QCD one postulates that the theory has $N$ different vacua as depicted in figure \ref{fig:thetapotential}.
At any given $\theta$ the vacuum energy is lowest in one of the $N$ vacua. As $\theta$ increases one finds that whenever it reaches an odd multiple of $\pi$ two of the vacua are degenerate. Further increasing $\theta$ past this point triggers a transition to a different vacuum and by the time we shifted $\theta$ by $2 \pi$ one indeed is back to the same physics, but in a different vacuum.

This allows a very simple description of the shallow interface, where the gradient of $\theta$ is small compared to the strong coupling scale of the theory. All the interesting physics is localized at the points where $\theta$ passes odd multiples of $\pi$. Since the gradient of $\theta$ is small, these loci are widely separated and we expect the total topological field theory to be simply $n$ copies of the theory living on a single interface, which then can be argued to be $SU(N)_{-1}$ based on anomalies \cite{Acharya:2001dz,Dierigl:2014xta}. For a steep interface there is another theory that can carry the correct anomaly -- a single $SU(N)_{-n}$. If the transition between the steep and shallow interfaces is second order, it should be described by a conformal field theory. This is the quiver gauge theory of  \eqref{eq:su_quiver_ym}. If we give all the scalars a large positive mass, we simply remove all bifundamental scalars leaving behind the product gauge group of the shallow interface. On the other hand, giving all the scalars a large negative mass drives the gauge groups into the Higgs phase, breaking them down to the single gauge group of the steep interface. We wish to see how this picture changes with the addition fundamental fermions.\footnote{For a discussion on interfaces in QCD$_4$, see Appendix A of \cite{Gaiotto:2017tne}.} This will require the use of the flavor violated master duality developed in the previous section.

To construct the flavorless quivers, we used the fact that one can identify and gauge the $SU(N_f) \times SU(N_s)$ flavor symmetries of (multiple copies of) the master duality to get bifundamental matter charged under various gauge symmetries. Adding flavors to the nodes can be achieved if we only gauge a \emph{part} of the flavor symmetry instead of the entire global flavor symmetry. The leftover global symmetry and the corresponding field components then become additional matter on each node.

We employ the aforementioned procedure in order to engineer a quiver with $SU$ gauge theories and scalar matter on the links. Node-by-node duality in general turns this into an equal length quiver with $U$ gauge theories on the nodes and once again scalar bifundamental matter. We add extra fundamental representation fermions on the original $SU$ side, which we will see necessitate extra scalar matter on the dual $U$ side. Intermediate steps involve theories with bifundamental fermions as well, but they are present neither in the initial nor in the final theory. The $U$ side can be argued to collapse to a single gauge group when the ranks of the gauge groups on all nodes are equal in the original $SU$ type theory. This pattern is what we previously found in the case of the unflavored quivers, and we will see it again once flavors are included.

While we can construct these quiver dualities for generic ranks and levels on the nodes (subject to certain bounds) there are special values of these parameters for which certain nodes confine. While in this case uncontrolled strong coupling dynamics is important, holography suggests that in this case a much simpler duality emerges. We will argue that
\begin{equation}
\label{eq:dualsu}
[SU(N)_{-1+ N_F/2}]^n + \text{bifundamental scalars} + \text{$N_F$ fundamental fermions per site}
\end{equation}
with certain potential terms is dual to
\begin{equation}
\label{eq:dualu}
 U(n)_{N} + \text{adjoint scalar} + \text{$N_F$ fundamental scalars}.
\end{equation}
Clearly this is a intuitive generalization of \eqref{eq:su_quiver_ym} and \eqref{eq:u_quiver_ym}. Let us justify these results again using node-by-node duality and holography.

\subsection{Node by Node Duality}
\label{subsec:Fermions-+su flq}

Before deriving the dual description for flavored quivers, let's remind ourselves of some notation. We will index the nodes by $i=1,\ldots,n$. The bifundamental scalars between the $i$th and $(i+1)$th node in the $SU$ and $U$ theories are labeled by $Y_{i,i+1}$ and $X_{i,i+1}$, respectively. The bifundamental fermions of the intermediate theories will be labeled by $\psi_{i,i+1}$. We'll call the flavored fermions belonging to the $i$th node $\psi_{i}$. Meanwhile, the scalar flavor degrees of freedom which are the dualized $\psi_{i}$ are denoted by $\phi_{i}$. All levels used in these notes are equivalent to the ``bare'' levels used in \cite{Komargodski:2017keh}.

\begin{figure}
\begin{centering}
\includegraphics[scale=0.7]{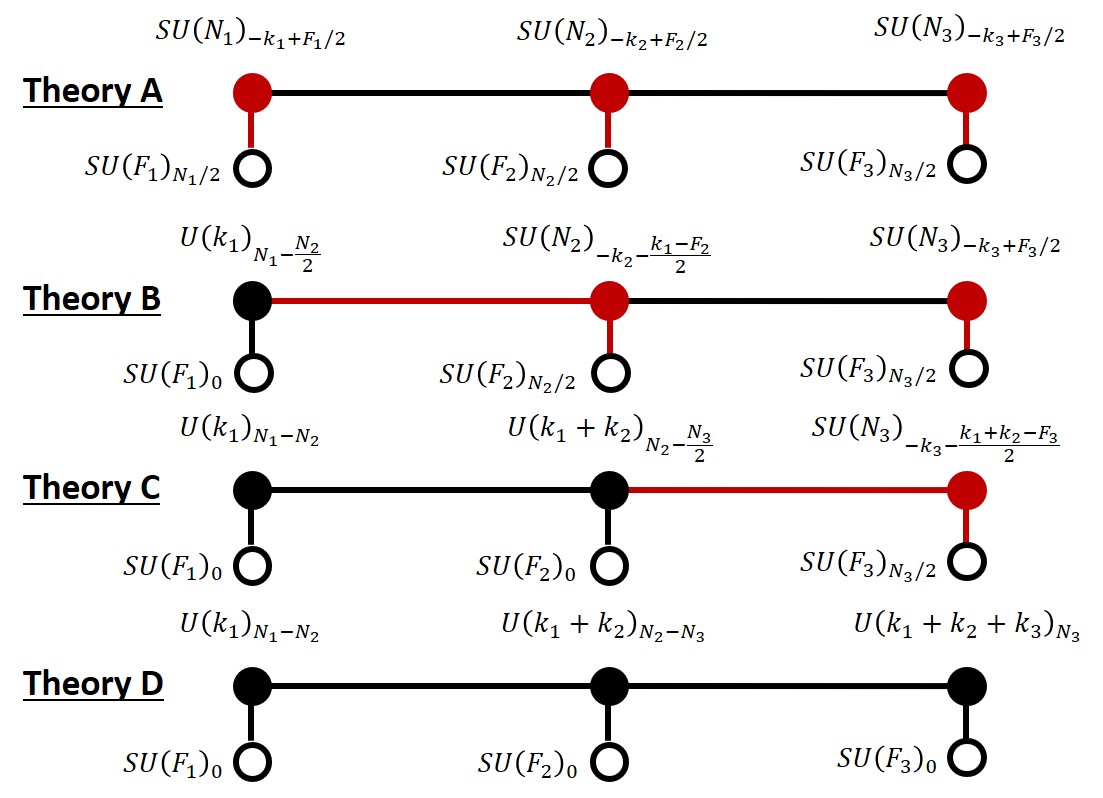}
\par\end{centering}
\caption{Flavored quiver with fermions on the $SU$ side and scalars on the $U$ side. The red/black nodes correspond to $U$ and $SU$ gauge groups, respectively. The red and black links are fermions and scalars. The white nodes are background flavor symmetries.\label{fig:ferm+su flq}}
\end{figure}

The most general 3-node quiver with flavor degrees of freedom on each node is shown in Fig. \ref{fig:ferm+su flq}. Its dual can be derived as in Ref. \cite{Aitken:2018cvh} using the master duality and its $N_f=0$ and $N_s=0$ limits.\footnote{In Ref. \cite{Aitken:2018cvh} the $N_F=0$ quiver was derived with no regard to distinguishing between ordinary and spin$_c$ connections. In Appendix \ref{sec:Quivers and spinc flq} we elaborate on how such quivers can be consistently formulated on spin$_c$ manifolds.} In particular, if one gauges the flavor symmetries of the master duality, one arrives at the following two and three-node dualities,
\begin{align}
SU(N)_{-k}\times \left[SU(N_s)_0\right]\qquad & \leftrightarrow\qquad U(k)_{N-N_{s}/2}\times \left[SU(N_s)_{-k/2}\right]\label{eq:dual1 flq}\\
SU(N)_{-k+N_{f}/2}\times \left[U(N_f)_{N/2}\right]\qquad & \leftrightarrow\qquad U(k)_N\times \left[U(N_f)_0\right]\label{eq:dual2 flq}\\
SU(N)_{-k+N_{f}/2}\times \left[U(N_f)_{N/2}\times SU(N_s)_0\right]\qquad & \leftrightarrow\qquad U(k)_{N-N_{s}/2}\times \left[U(N_f)_0\times SU(N_s)_{-k/2}\right].\label{eq:dual3 flq}
\end{align}
Each of these is subject to particular flavor bounds, but let us ignore them for a moment. Stepping from Theory A to Theory B we use \eqref{eq:dual3 flq} with the $U(N_{f})$ symmetry ungauged (i.e. the master duality with only the $SU(N_{s})$ symmetry promoted to be dynamical).
From Theory B and Theory C, again use \eqref{eq:dual3 flq} but gauge only part of the background flavor symmetry such that the $U\left(k_{1}+F_{2}\right)$ background fermion flavor symmetry becomes a $U\left(k_{1}\right)$ gauge symmetry and the remaining $SU\left(F_{2}\right)\times U\left(1\right)$ are still global symmetries.
Finally, use \eqref{eq:dual2 flq} to go from Theory C to Theory D, again with a split flavor symmetry with a part which is gauged and another which is untouched.

It is straightforward to see how this pattern generalize to the $n$-node quiver. Let $N_1$, $\ldots$, $N_n$ denote the number of colors on the nodes and $k_1$, $\ldots$, $k_n$ the levels of the corresponding Chern-Simons terms. The duality for the $n$-node quiver reads
\begin{subequations}
\label{eq:general_n}
\begin{align}
\left[ SU(N_{1})_{-k_{1} +\frac{F_1}{2}} + \text{fund } \psi_1 \right ] \times\prod_{i=2}^{n}\left[SU(N_{i})_{-k_{i}+\frac{F_i}{2}}+\text{bifund }Y_{i-1,i} + \text{fund } \psi_i \right]\\
\leftrightarrow \qquad \prod_{i=1}^{n-1}\left[U(K_{i})_{N_{i}-N_{i+1}}+\text{bifund }X_{i-1,i}
+ \text{fund } \phi_i \right]\times \left [ U(K_{n})_{N_{n}} + \text{fund } \phi_n \right ].
\end{align}
\end{subequations}
The ranks of the gauge groups in the $U$ quiver are given by
\begin{equation}
K_n \equiv \sum_{i=1}^n k_i.
\end{equation}
This is very similar to the relation as was found in the unflavored case \cite{Aitken:2018cvh}. As in the 3-node case depicted in Fig. \ref{fig:ferm+su flq} we can also keep track of the level of the flavor groups. These can always be shifted by an overall background Chern-Simons term added on both sides, but if we chose the levels to be $SU(F_i)_{N_i/2}$ on the $SU$ side, they end up being $SU(F_i)_0$ on the $U$ side.

Now let us specialize to the case where $N_{1}=\ldots=N_{n}=N$, $k_{1}=\ldots=k_{n}=1$, and $F_{1}=\ldots=F_{n}=N_{F}$. If, in addition, we add a $\det X_{i,i+1}$ potential term on each link, this is the theory that describes the physics of interfaces in QCD$_4$. Recall, the determinant term was needed in order to eliminate the additional $U(1)^{n-1}$ global symmetry present on the quivers \cite{Aitken:2018cvh, Gaiotto:2017tne}, and its $U$ side equivalent is a monopole operator. It is easy to see that only the $N_{F}=0$ case does not violate any flavor bounds. This is because all the levels must satisfy $k_{i}\geq F_{i}$ and in addition we must avoid the double-saturation limit of the master duality. Since we have already saturated the $N_{i}\geq N_{i+1}$ flavor bound, $N_{F}=0$ is the only way we can avoid violating such bounds.

The flavor-extended master duality and double saturated master we laid out in the previous section allows us to proceed. The $N_F=1$ case corresponds to use of the double-saturated master duality. As discussed in Sec. \ref{subsec:double-sat}, the introduction of the determinant terms allow extend the validity of the master duality to this limit.  We will discuss the $N_F>1$ in detail in what follows below, but let us summarize here the essential points for general $N_F\geq 1$. We saw that for one sign of the mass the flavor extended duality is just the same as we had when the flavor bounds were obeyed (see \eqref{eq:fvmd 3} and \eqref{eq:fvmd 4}) , so we can just continue to use the duality above if we stay in such phases. On the $SU$ side, each of the nodes has a
\begin{equation}
SU(N)_{-1+N_{F}/2}\text{ with \ensuremath{N_{F}} \ensuremath{\psi}}
\end{equation}
theory on it. We also see that in this special case of equal $N_i$ we get a dual quiver on the $U$ side where all but the last node have a level 0 Chern-Simons term. This is once again completely identical to the case of the un-flavored quivers considered in \cite{Aitken:2018cvh}. The non-Abelian factors with level 0 confine. Lo and behold, all but the last node disappear from the low energy spectrum. To pin down the remaining charged matter under the last gauge group, we need to make some dynamical assumptions for this confinement mechanism. In \cite{Aitken:2018cvh} it was argued that the only light remnant of the bifundamental matter is an adjoint ``meson" made from the bifundamentals on the last link. This assumption gave rise to a duality conjecture that agreed with the holographic construction and passed several non-trivial consistency tests. If we make the same assumption here, we are lead to the duality conjecture of eqs. (\ref{eq:dualsu}) and (\ref{eq:dualu}). Once again, we will see that this is also what the holographic construction tells us.

Similar dualities can also be derived if the extra flavors on the $SU$ side are taken to be scalars, but the patterns that emerge are more complicated than in the fermionic case and do not appear to yield any simple interesting new dualities.

%

\subsubsection{Constructing the $N_{F}>1$ Quiver}

Let us now use the flavor extended master duality to construct quivers. This will allow us to extend the validity of our dualities to $N_{F}>1$. As we have seen, this will be complicated slightly by the fact one needs two theories to describe the entire phase diagram of the $U$ side. Since there are even more mass deformed phases in the quivers, we should expect the number of theories needed to ultimately describe all possible mass deformations of the $U$ side to be quite large. Every time one uses a flavor violated duality the number of scalar theories needed to describe the entire phase diagram doubles. For our three-node case we can consider $2^{3}=8$ separate theories to capture all possible mass deformations of the original fermionic theory. However, all eight theories should be equivalent descriptions of the same Grassmannian manifold.

In this section we will only consider two of the eight possible descriptions of the $U$ side of the quiver. These will be the cases which are valid when all masses of the quivers are deformed such that we are in phases III and IV or I and II  of the flavor-extended master duality. These correspond to the extreme cases in the original $SU$ theory where all the fermions are tuned to large positive and large negative mass, respectively. It is possible construct quivers for mixed phases, but we do not consider them here.


In what follows, we will first apply this general strategy to the special case of $n=3$ nodes. After considering the three-node case, we will generalize such extreme
cases to $n$-nodes.

\subsubsection*{$m_{\psi_i}<0$ Duality}

The special case where all masses are negative is especially easy since in this case the duality map is formally still given by (\ref{eq:dual3 flq}).  That is, the entire derivation shown in Fig. \ref{fig:ferm+su flq} in principle still holds since the initial theory is the same and the extended master/Aharony dualities are also the same. Thus much of the same conclusions we made there can be applied here. The primary difference is that the identification between the two sides of the phase diagrams is shifted from the flavorless case. Following the mass mappings through the derivation, for the bifundamentals we still have $m_{Y_{1,2}}^{2}\leftrightarrow m_{\psi_{1,2}}^{2}\leftrightarrow-m_{X_{1,2}}^{2}$. For the flavor degrees of freedom we have
\begin{equation}
m_{\psi_{i}}+m_{*}\leftrightarrow-m_{\phi}^{2}.
\end{equation}
Zero mass for the fermion no longer maps to zero mass for the boson. In particular, the $m_{\psi_i}=0$ region now corresponds to the $m_{\Phi_i}^{2}<0$ region, and this ``mass offset'' will introduce certain complications.

Let us discuss in detail how the analysis of interfaces changes in the presence of the additional flavors due to this mass offset. Once more we specialize to the values relevant for QCD$_4$, namely $N_i=N$, $k_i=1$, and $F_i=N_F>1$. We'll begin by assuming no deformations of the matter on the $SU$ side, i.e. $m_{\psi_{i}}=0$ and $m_{X_{i,i+1}}^{2}=0$. First off, note that because $N_{F}>1$ each node in the initial $SU$ theory has Grassmannians. Thus, this should ultimately map to something on the $U$ side of the duality which also produces Grassmannians. Since the $U$ side only has scalars, in order to produce said Grassmannians some of said scalars must be in their Higgs phase. This is already very different than the flavorless case where theories with no mass deformations were mapped to one another.

It will be useful to keep in mind which matter is ``responsible'' for the flavor violation and presence of the Grassmannian description. Recall, per the conjectured of \cite{Komargodski:2017keh}, this can occur from either having too many light fermions or too many scalars with a negative mass. The culprit is obvious in the $SU$ theory: without the additional flavors on each node we have the flavorless case which we know meets all flavor bounds. Thus, it is the additional fermions of each node which are responsible.

This is complicated in (say) Theory B, where we now have two subgroups of fermions connected to the second node, corresponding to $SU(F_2)$ and $SU(k_1)$ (see Fig. \ref{fig:ferm+su flq}). The former are the additional flavors that were on that node to begin with, the latter are the bifundamental fermions which were previously scalars. If all such fermions were light for the portion of the phase diagrams we are interested in, the Grassmannian manifold of the second node may have changed, and this would be problematic since we claim it is dual to Theory A.

To resolve this issue, first note that in Theory B the flavor degrees of freedom on the first node have been converted into scalars, and because the mass offset these scalars have a negative mass. This is consistent with there being a Grassmannian on the first node. What we have neglected are the interactions introduced by using the master duality that are now present in Theory B. Since the scalar flavors on the first node are now Higgs'd, the interactions cause the $\psi_{1,2}$ fermions to be gapped. Meanwhile, the $\psi_{2}$ remain light. Hence the only light fermionic degrees of freedom are those which were already present in Theory A, so the Grassmannian on the second node also remains the same.

Generalizing this argument down the quiver, we see that it is always either the $\psi_{i}$ or $\phi_{i}$ (i.e. the flavor degrees of freedom on each node) which break the nodes down to Grassmannian descriptions. This might have been expected, since these are the new ingredients relative to the flavorless case, but it is nice that the quivers get it right.

As in Ref. \cite{Aitken:2018cvh}, the phase corresponding to $m_{X_{i,i+1}}^{2}<0$ causes a breaking of the Chern-Simons terms in the theory to $\left[U(1)_{N}\right]^{3}$. Thus on the $U$ side of the theory we ultimately end up with
\begin{equation}
U(1)_{N}\text{ with \ensuremath{N_{F}} \ensuremath{\phi_{i}}}
\end{equation}
on each node, with $m_{\phi_{i}}^{2}<0$, as we would have expected.

\paragraph{Critical Theory}
Formally the duality map \eqref{flavor_violated_mast} is only valid when we tune the fermion mass to it's critical value, the scalar mass to zero and apply the map \eqref{eq:neg_mass_map}. In this case the dualized scalars $\phi_{i}$ are massless and the interactions terms introduced through the master duality do nothing to the mass of the fermions. This is okay since we would not expect Grassmannians at this location in phase space. Additionally, since the number of bifundamental fermions is less than the level of the corresponding node, we do not get any Grassmannians. The duality thus states
\begin{multline}
\left[SU(N)_{-1}+N_{F}\text{ fundamental fermions with \ensuremath{m_{\psi_{i}}=-m_{*}}}\right]^{n}+\text{bifundamental scalars}\\
\leftrightarrow\qquad U(n)_{N}+N_{F}\text{ fundamental scalars}+\text{bifundamental scalars}.
\end{multline}

It is interesting to consider deformations of the bifundamentals in this phase. Luckily, almost all of the analysis performed in \cite{Aitken:2018cvh} holds here as well. That is, on the $SU$ side gapping the bifundamentals simply removes them from the spectrum. Higgsing the bifundamentals leaves the gauge transformations which transform two adjacent nodes equally unbroken. This causes the levels of the two nodes to add. The mass identifications of the bifundamentals on the $U$ side are still opposite of those on the $SU$ side. Thus for gapped scalars on the $SU$ side we have Higgsing of the $U$ groups down to $U(1)$. Meanwhile, the first $n-1$ nodes on the $U$ side are confining for full Higgsing on the $SU$ side.

Let us consider this latter case in a little more detail. Consider sitting at the critical point where all $m_{\psi_{i}}=-m_{*}$ and all $m_{Y_{i,i+1}}^{2}=0$. Now give all $Y_{i,i+1}$ a negative mass, so the Chern-Simons term on the $SU$ side is $SU(N)_{-n}$. All bifundamentals are gapped on the $U$ side, and the only non-confining node is all the way on the right, given by a $U(n)_{N}$ Chern-Simons theory. Now consider tuning the masses of the flavor degrees of freedom. Move the fermion on the last node, $\psi_{n}$, to a slightly larger mass (but smaller in magnitude), $-m_{*}+\epsilon$, which causes us to get a $\mathcal{M}(N_F,n)$ Grassmannian. On the $U$ side this causes the dual scalar $\phi_{n}$ to acquire a negative mass, breaking down the node flavor symmetry to $\mathcal{M}(N_{F},n)$. Thus both sides are consistent.

One needs to be slightly more careful when tuning the masses of any other matter, say $\psi_{j}$ for $j<n$. Moving to a larger mass $m_{\psi_{j}}=-m_{*}+\epsilon$ again causes the corresponding $\phi_{j}$ scalar to get a negative mass, which due to interaction terms present on the $U$ side gives the $X_{j,j+1}$ bifundamental a \emph{negative }mass. Although the interactions are finite, we will only focus on the case when they are stronger than mass deformations. The scalar then breaks the corresponding the $U(j+1)_{0}$ gauge group and this propagates down the quiver similar to the flavorless case. Ultimately this causes a breaking of $U(n)_{N}\to U(n-j)_{N}\times U(j)_{N}$. On the $SU$ side of things this is reflected in the fermion condensing, causing a breaking of the gauge group and a $\mathcal{M}(N_{f},j)$ Grassmannian on the $j$th node, and a leftover $SU(N)_{-n+j}$. This is consistent since the breaking from the negative mass bifundamental scalar, $Y_{j,j+1}$, is inconsequential for corresponding node since it is instead now broken by the fermion condensate. To summarize, tuning the masses of flavor degrees of freedom on nodes $j<n$ undoes the effects of Higgsing the bifundamental scalars on the $SU$ side/gapping
the bifundamental scalars on the $U$ side.

\subsubsection*{$m_{\psi_i}>0$ Duality}

\begin{figure}
\begin{centering}
\includegraphics[scale=0.7]{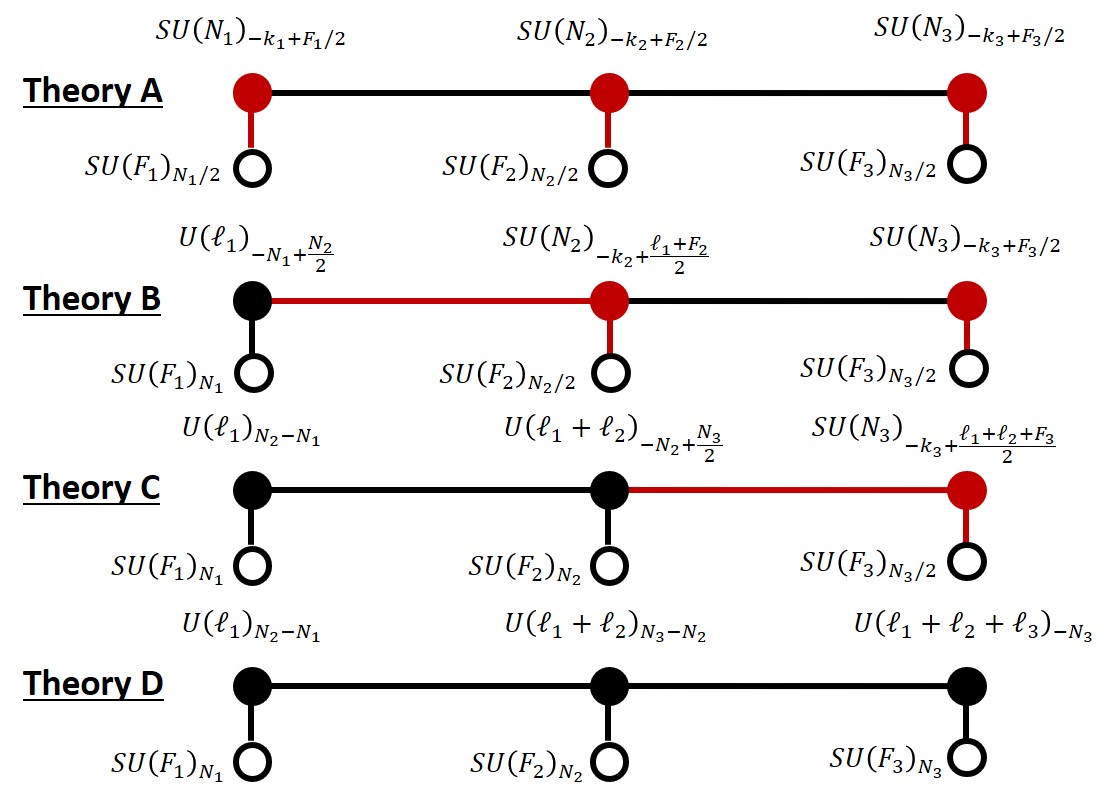}
\par\end{centering}
\caption{Flavored quiver with fermions on the $SU$ side and scalars on the $U$ side. This is an alternate theory in the flavor extended case. Here we have introduced the notation $\ell_i\equiv F_i - k_i$. \label{fig:ferm+su alt flq}}
\end{figure}

Now let us instead look at the case where all masses are positive. In this case we need to use the $m_\psi = m_*$ theory of \eqref{flavor_violated_mast}, which is spelled out explicitly in \eqref{eq:master ext pos flq}, to derive the flavor extended quiver. We consider again the three node quiver for the most general set of parameters. It will be useful to introduce a new notation $\ell_{i}\equiv F_{i}-k_{i}$. The derivation is shown in Fig. \ref{fig:ferm+su alt flq}.

Note since we are using the $m_\psi = m_*$ duality, the interaction in this phase is slightly different. Namely, the interactions are now in the $m_{\Phi}^{2}<0$ and $m_{\Psi}>0$ phase and prevent the fermions from getting a positive mass. That means that the interaction term comes with a negative sign out front.

\subsection{Holographic Construction}

Having generalized the node-by-node construction in order to motivate the duality between \eqref{eq:dualsu} and \eqref{eq:dualu}, let us now turn to the holographic derivation. The holographic construction used in  \cite{Aitken:2018cvh} is in spirit similar to the one first introduced in \cite{Jensen:2017xbs} for the study of 2+1 dimensional bosonization dualities, which is based on the even earlier holographic realization of level/rank duality in \cite{Fujita:2009kw}. The key idea here is to embed the Chern-Simons matter theory of interest inside a larger gauge theory with a known holographic dual. In the UV we have the full set of degrees of freedom of the larger gauge theory, dual to a theory of gravity in the bulk. Upon suitable deformation on both sides the gauge theory and its gravitational dual develop a gap for most degrees of freedom, only a small subset survives in the infrared. The original field theory is engineered to reduce to a Chern-Simons matter theory in 2+1 dimensions. In the bulk most degrees of freedom are gapped out as well, including all the fluctuations of the graviton and its superpartners. The only degrees of freedom that survive are localized on a probe D-brane.

More specifically, in \cite{Jensen:2017xbs} the basic bosonization dualities were reproduced by compactifying ${\cal N}=4$ super Yang-Mills on a circle with antiperiodic boundary conditions for the fermions and a $\theta$ angle wrapping non-trivially around the circle direction. The low energy theory yielded a gapped Chern-Simons theory, with extra light matter added via probe branes. In order to obtain quiver gauge theories one needs to start with a slightly more complicated construction. Fortunately, most of this framework has been laid out in \cite{Witten:1998uka} in the context of the simplest holographic dual for a confining gauge theory: Witten's black hole \cite{Witten:1998zw}. Witten's black hole describes the gravitational dual of $4+1$ dimensional super Yang-Mills compactified on a circle with anti-periodic boundary conditions for the fermions. This procedure gives masses to all matter fields and so, at low energies, one is left with pure Yang-Mills in $3+1$ dimensions, which itself is gapped. Topologically Witten's black hole solution is $\mathbb{R}^{3,1} \times D_2 \times S^4$, where the first factor is $3+1$ dimensional Minkowski space and the disc or ``cigar" $D_2$ contains the radial direction of the holographic dual as well as a compact circle direction which smoothly shrinks to zero size at a critical value $r_*$ of the radial coordinate. The relevant holographic dual for the interfaces we are seeking are D6 branes sitting at the bottom of the cigar, wrapping all of the internal $S^4$ and being localized in one of the three spatial directions of the $\mathbb{R}^{3,1}$ factor.

One subtlety here is that the D6 branes were argued in \cite{Witten:1998uka} to be dual to a domain walls between two vacua at a fixed $\theta$ rather than a $\theta$ interface. At a given $\theta$ we still have $N$ vacua, with all but one of them being only meta-stable. The energy difference between vacua however is of order 1 in the large $N$ counting, whereas the tension of the domain wall is of order $N$. So in the large $N$ limit these domain walls are long lived and approximately stationary. To truly describe an interface, we should turn on a $\theta$ gradient in addition to the D6 branes, which corresponds to a RR 1-form in the bulk that depends on the radial direction in the bulk as well as the spatial direction orthogonal to the domain walls,  call it $x$. The simplest such supergravity solution is an RR 1-form that grows linearly along $x$. In this case the RR 1-form equation of motion is solved without giving the 1-form any dependence on the holographic radial coordinate. This solution as it stands corresponds to increasing $\theta$ along a spatial direction without adjusting vacua, but instead staying on a meta-stable branch. This is clearly not the physical vacuum of the theory. In order to truly correspond to the interfaces of \cite{Gaiotto:2017tne} we need to still include D6 branes in addition to the spatially varying theta angle in order to ensure that the field theory always is living in the locally true vacuum. At least in the simple case of a linearly varying $\theta$ angle it is easy to see that the D6 branes experience a potential that will automatically localize them where $\theta$ crosses odd multiples of $\pi$. To see that this is the case, recall that the backreaction of the D6 branes induces an RR 1-form field $\sim (\theta + 2 \pi K)$ which gives rise to an energy density of order $(\theta + 2 \pi K)^2$. Here $K$ labels the different vacua. Let us look at a spatial region in which $\theta$ varies over a $2 \pi$ range. Let us chose this range to be 0 to $2 \pi$ with $K=0$. This can always be done by the way we label vacua in terms of $K$. Let us assume that the change is $\theta$ is taking place over an interval of length $L$ with $\theta = 2 \pi x/L$. There is a single D6 brane located at $x_0$ with $0 \leq x_0 \leq L$ across which the vacuum jumps to $K=-1$. The vacuum energy associated to this configuration is
\beq
 E \sim \int_0^l dx \, (\frac{2 \pi}{L} x - \Theta(x-x_0) 2 \pi)^2  = \frac{4 \pi^2(L^2 -3 L x_0 + 3 x_0^2)}{3L}
 \eeq
where $\Theta$ is the step function that is 1 when its argument is positive and 0 otherwise.
Minimizing with respect to $x_0$ we indeed find $x_0=L/2$. The D6 brane wants to sit at the middle of the interval where $\theta=\pi$. Note that these energies are all of order 1 in the large $N$ counting. This is due to the fact that they involve the brane backreaction. When determining the leading order in $N$ physics on the brane, we can neglect these potentials. The upshot is that irrespective on whether we describe interfaces or domain walls, the holographic dual description is given in terms of a stack of $n$ D6 branes at the bottom of the cigar as long as we jump $n$ vacua across the co-dimension one object.

Interestingly, this holographic dual at low energies does not give back the quiver gauge theory \eqref{eq:su_quiver_ym}, but its dual incarnation \eqref{eq:u_quiver_ym}: The theory on a stack of $n$ D6 branes is a $U(n)$ super-symmetric gauge theory. Due to the Wess-Zumino terms on the worldvolume coupling to the 4-form flux supporting Witten's black hole the gauge field picks up a Chern-Simons term of level $N$. The worldvolume fermions get mass from the compactification on the internal $S^4$. The worldvolume scalars correspond to geometric fluctuations of the D6 branes. The two scalars corresponding to fluctuations of the D6 on the disc are massive due to the warped geometry of the spacetime: there is a non-vanishing potential energy cost associated with fluctuating up the cigar. The only light matter on the D6 branes is a single massless adjoint scalar corresponding to the fluctuations in the $x$ directions. To leading order in $N$ it is massless, even though we've already seen above that a non-trivial potential will surely be generated at order 1. Lo and behold, the gauge theory on the D6 branes exactly realizes \eqref{eq:u_quiver_ym} as advertised.

The inclusion of extra flavors is now conceptually straightforward but the details are somewhat daunting. Many aspects of this construction have been discussed nicely in \cite{Argurio:2018uup}. On the field theory side, our flavored quiver gauge theory from \eqref{eq:dualsu} indeed already appeared in the study of interfaces alongside its flavorless cousin. If instead of studying $\theta$ interfaces in pure Yang-Mills one studies them in a confining $SU(N)$ gauge theory with fundamental fermions, the correct gauge theory on the domain walls is exactly the flavored quiver \cite{Gaiotto:2017tne}. In the holographic dual, we need to augment Witten's black hole with $N_F$ flavor D8 branes in order to describe holographic QCD \cite{Sakai:2004cn}. The D8 branes are localized on the compact circle but their worldvolume extends in all other directions. In particular, the D6 branes dual to the domain walls are entirely embedded inside the D8 worldvolume. That is, the D6/D8 system constitutes a Dp/Dp+2 brane system with 2ND directions. Correspondingly, the 6-8 strings connecting the two gives rise to extra scalar matter in the fundamental representation of the $U(n)_N$ gauge theory on the D6 worldvolume. Indeed we have found that the flavored quiver of \eqref{eq:dualsu} has a holographic dual description in terms of the theory in \eqref{eq:dualu}. When asking more detailed questions, lots of open problems emerge. The scalar at a 2ND brane intersection is tachyonic. So in order to find the CFT dual to the flavored quiver, we need to let the scalars condense. The result is outside of the range of perturbative string theory, so the outcome is somewhat inspired guesswork. The fact that this condensation is happening is presumably related to the fact that in the field theory we passed the flavor bound. The extension of the duality into this regime moved occurrence of the conformal field theory in the phase diagram from zero mass to the edge of the chirally broken regime with its Grassmannian pion Lagrangian as already argued in \cite{Argurio:2018uup}. So qualitatively holography supports our duality conjecture.

\subsection{Phase Matching}

As a final check we want to confirm that the possible phases match in the duality of (\ref{eq:dualsu}) and (\ref{eq:dualu}). Recall how the matching of phases worked in the flavor-less case. On the $SU$ side the bifundamental scalars could either acquire an expectation value or a mass. In the latter case they just disappear from the low energy spectrum, in the former they break two neighboring nodes to the diagonal subgroup. The corresponding levels add. The allowed phases were hence given by partitions $\{ n _I \}$ of $n$, that is integers $n_I$ with $\sum_I n_I = n$. Each $n_I$ specified how many consecutive nodes were broken down to the diagonal subgroup before encountering a scalar that became massive. So $n_1=n$ corresponds to the case were all bifundamentals acquired an expectation value, whereas $n_I=1$ for $I=1,\ldots, n$ is the case where all bifundamentals get a mass. The generic partition led to a phase governed by a topological field theory based on a gauge group
\begin{equation} \label{eq:suphases} \prod_I SU(N)_{- n_I}. \end{equation}
The phases of the $U(N)_n$ gauge theory were parametrized by the expectation values of the adjoint scalar. Latter can always be diagonalized, so we have to specify $N$ eigenvalues. Enhanced unbroken gauge groups arise whenever eigenvalues coincide. Using again the partition $\{ n_I \}$ to denote to multiplicities of repeated eigenvalues the phases of the $U(N)_n$ plus adjoint theory gave
\begin{equation} \label{eq:uphases} \prod_I U(n_I)_{N}. 
\end{equation}
Eqs. \eqref{eq:suphases} and \eqref{eq:uphases} are level-rank duals, so both sides have the same phase diagram.

Now let us see what happens in the presence of flavors. Let us first take a look at the $U$ side of the duality. Under the gauge symmetry breaking of \eqref{eq:uphases} triggered by the adjoint scalar the fundamental matter multiplets decompose into fundamental matter under each of the product factors, so that we obtain a theory
\begin{equation} \label{eq:uphasesflavor} \prod_I \left [ U(n_I)_{N} + N_F \text{ fund scalars} \right ]. 
\end{equation}
For each gauge group factor we now can, as usual, drive the fundamental scalars to condense or to become heavy and decouple. We know that the dual description of all these phases is captured by:
\begin{equation} \label{eq:suphasesflavor} \prod_I \left [ SU(N)_{- n_I + N_F/2} + N_F \text{ fund fermions} \right ]. \end{equation}
This is indeed a theory we can get out of the flavored quiver, but it requires a non-trivial potential. Note that in the quiver without potential we would find that whenever two nodes with their $SU(N)_{-1+N_F/2}$ gauge groups and $N_F$ fundamental fermions each get broken down to their diagonal subgroup, we would get a $SU(N)_{-2 + N_F}$ and $2 N_F$ fundamental fermions and an enhanced flavor symmetry. In order to get  \eqref{eq:suphasesflavor} we need a quartic potential which gives one of the two sets of $N_F$ flavors a negative mass so they decouple and shift the Chern-Simons level back to $N_F/2$. Generally, when we break down $n_I$ nodes to their diagonal factor, we need $n_I-1$ sets of $N_F$ flavors to get a negative mass from the $n_I-1$ bifundamentals. This can be accomplished by the appropriate quartic potential, but it is crucial that this is included. Apparently the limit of no interactions on the $SU$ side that we have mostly been working with in the flavorless case is inconsistent with the confinement scenario where we are left with the theory of \eqref{eq:dualu} on the $U$ side.

\subsection{Enhanced Flavor Symmetries}
\begin{figure}
\begin{centering}
\includegraphics[scale=0.6]{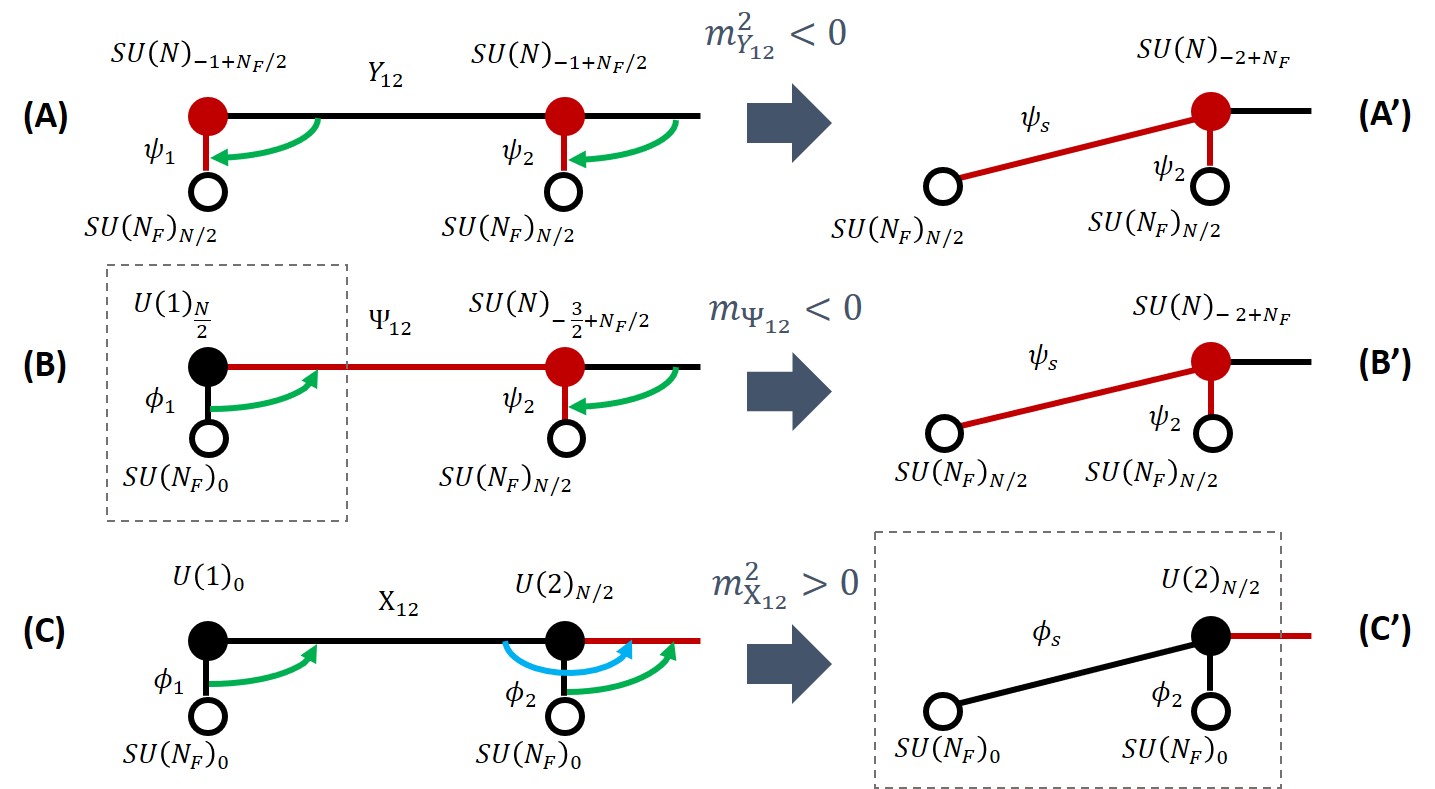}
\par\end{centering}
\caption{Explanation of interaction and enhanced flavor symmetries which arise due to the flavor-violated master duality and its finite interaction terms. We are using green/blue arrows to denote the (unidirectional) interactions. The green arrows represent finite interactions between the flavor degrees of freedom and adjacent bifundamentals. The light blue arrows represent interactions between bifundamental scalars, which were also present in the pure YM case of \cite{Aitken:2018cvh}. We assume they take the very same form they did there. The right-hand side shows how one arrives at enhanced flavor symmetries on the adjacent node at each step of the duality.\label{fig:enhanced_flavor_Sym}}
\end{figure}

A crucial feature of the new flavor-violated master duality is the presence of finite interactions on the $SU$ side of the duality. These interactions ensure that when one Higgses the bifundamental scalars they give a gap to all but the very last node's flavor degrees of freedom. Explicitly, the interactions enter in the form\footnote{For consistency we must also include finite interactions for scalars on adjacent links, however we are only focusing on the part of phase space where these interactions are small compared to any given mass scalar mass deformation. We leave a full mapping of the phase diagram with finite interactions for future work.}
\begin{equation}
\mathcal{L}\subset c^{\prime}\sum_{i=1}^{n-1}\left(Y_{i,i+1}^{\dagger}\psi_{i}\right)\left(\bar{\psi}_{i}Y_{i,i+1}\right).
\end{equation}
Thus, when we Higgs the $Y_{i,i+1}$ bifundamentals, the $\psi_{j}$ for $j=1,\ldots,n-1$ all acquire a finite positive mass. Since only the $\psi_{n}$ remains light, we effectively have an $SU(N)_{-n+N_{F}/2}$ theory coupled to only $N_{F}$ fermions. If these interaction terms were not present, all $\psi_{i}$ would remain light. This would lead to an $SU(N)_{-n+N_{F}/2}$ theory coupled to $nN_{F}$ fermion, which gives a Grassmannian which does not agree with the pure $3+1$ dimensional analysis \cite{Gaiotto:2017tne}.  

Although the $\psi_{j}$ acquire a mass from interaction terms, this can be canceled by an \textit{explicit} mass deformation for all the $\psi_j$. This is what occurs on critical lines corresponding to II-III, II-VI, and III-VI. If this is done for all $\psi_{j}$, then we once more  have $nN_{F}$ light fermions, which can lead to enhanced symmetries and strange looking Grassmannians. For consistency, these flavor enhanced Grassmannians should be present on the $U$ side of the duality as well. This is indeed what occurs. The procedure is summarized in Fig. \ref{fig:enhanced_flavor_Sym}.

First let us review what occurs in the fully $SU$ theory, denoted by A in Fig. \ref{fig:enhanced_flavor_Sym}. When one Higgses the $Y_{12}$ bifundamental, this effectively ties the gauge theories of the first and second node together. We have denoted this mass deformation in A' by representing said nodes as a single node. From the point of view of Fig. \ref{fig: master_fv_n=ns}, this corresponds to moving along the II-III critical line. Now, when one gauges the $SU(N_s)$ global symmetry, one gets a Grassmannian around where this line was previously. This is because the non-Abelian background terms have Chern-Simons levels which violate the flavor bounds. This Grassmannian will be $\mathcal{M}\left(2N_{F},2\right)$, where the $2N_{F}$ comes from the $N_{F}$ fermion flavors which were already tied to the second node, and another $N_{F}$ flavors which previously belonged to the first node.

Now let us discuss how this is matched in B of Fig. \ref{fig:enhanced_flavor_Sym}. To do so, it is helpful to reference what occurs in Fig. \ref{fig: master_fv_n=ns} along the II-III critical line when the $SU(N_s)$ global symmetry is gauged. Note that, prior to the gauging of said symmetry, we also get $N_{s}N_{f}$ light fermions along said line, which we will call $\psi_{s}$ \footnote{As a quick aside, notice that it is difficult to see from the point of view of just the blue scalar theory where these light fermions come from (as opposed to the III-VI critical line, where all $kN_{s}$ fermions modes are simply light). Said light fermions are needed to explain the change in background Chern-Simons terms between phases II-III, which gives us confidence their number is correct. Abstractly, they can be viewed as the two critical lines II-VI and III-VI merging together, but these singlet fermions are described by the two distinct scalar theories (i.e. the red and blue ones).}. 
From Fig. \ref{fig: master_fv_n=ns}, we get the same behavior when we give $\Psi_{12}$ a negative mass deformation. The $\psi_{s}$ lead to an enhanced flavor symmetry on the second node, shown in B' of Fig. \ref{fig:enhanced_flavor_Sym}. If we took the $\psi_{s}$ and $\psi_{2}$ to be light, we would again arrive at the Grassmannian $\mathcal{M}(2N_{F},2)$. Everything seems consistent so far, which shouldn't be too surprising since we have only used the flavor-violated master duality once, and it can be confirmed this is consistent when gauging the $SU(N_{s})$ symmetry.

In moving from B to C we have used the flavor-violated master duality on the second node. This changes the fermions $\Psi_{12}$ and $\psi_{2}$ to scalars. Strictly looking at C and gapping the $X_{12}$ scalars, which via mass mapping is analogous to Higgsing the $Y_{12}$, it is difficult to see how we again get the enhanced flavor symmetry on the second node. This isn't too surprising, because even in moving from B to B' it was difficult to explain where the $N_{s}N_{f}$ $\psi_{s}$ came from. Instead, consider applying the master duality to B' (i.e. B with $\Psi_{12}$ already deformed to large negative mass). This would change the $N_{F}$ light $\psi_{2}$ and $N_{F}$ light $\psi_{s}$ to corresponding light scalars in C'. Thus from this point of view, it makes sense that we get $N_{F}$ light $\phi_{2}$, but also $N_{F}$ light $\phi_{s}$, leading once more to an enhanced flavor symmetry. Similar to $\psi_{s}$, the we can conjecture the $\phi_{s}$ are light by some cancellation between the interaction terms and the explicit mass deformations. Since we now have $2N_{F}$ light scalars, if we take them to negative mass we once more find the $\mathcal{M}(2N_{F},2)$ Grassmannian.

To see this continues to occur as we move to more general quivers,
note that we can continue to use these very same arguments used above
to see enhanced flavor symmetries in $U$ theories. In particular,
in Fig. \ref{fig:enhanced_flavor_Sym}, the part of the quiver in
the dashed box of C' is identical to that of B, except for the fact
the former has $2N_{F}$ light scalars instead of only $N_{F}$ light
scalars. Thus, repeating the very same arguments used in moving from
B to B' above, we come to the conclusion that as one takes $\Psi_{23}$
to negative mass, we should again get light $\psi_{s}$ fermions bifundamentally
charged under the scalar and fermion flavor symmetries. However, now
those symmetries are $SU\left(N_{s}\right)\times SU\left(2N_{F}\right)$.
This would then lead to an $SU\left(3N_{F}\right)$ symmetry on the
third node.

Lastly, let us propose where the extra light $\psi_{s}$ and $\phi_{s}$
we saw above in theories B' and C' arise. There are very natural bound
states which have the exact quantum numbers to match these bifundamentals,
namely
\begin{equation}
\psi_{s}=\phi_{1}\Psi_{12},\qquad\phi_{s}=\phi_{1}X_{12}.
\end{equation}
Furthermore, the corresponding $U$ node under which each component
of these bound states is charged (i.e. node one) is confining. Thus
we conjecture that some combination of confinement and mass deformations
somehow makes said bound states light. This procedure generalizes
as one moves down the quiver. For example, the additional light scalars
one gets on the third node for certain mass deformations could correspond
to bound states $\phi_{1}X_{12}X_{23}$ and $\phi_{2}X_{23}$. Along with
$\phi_{3}$, said scalars can lead to an enhanced flavor symmetry.

\section{Orthogonal and Symplectic Gauge Groups}
\label{sec:orth and symp}

\subsection{Node-by-node}
We can generalize the duality for quiver gauge theories to the case of symplectic and orthogonal gauge groups. The node-by-node construction generalizes in a straightforward fashion. According to \cite{Aharony:2016jvv} the basic bosonization dualities in this case are
\begin{subequations}
\begin{align}
SO(N)_k + N_f \mbox{ real scalars} & \qquad\leftrightarrow\qquad & SO(k)_{-N+N_f/2 } + N_f  \mbox{ real fermions} \\
Sp(N)_k + N_f \mbox{ real scalars} & \qquad\leftrightarrow\qquad & Sp(k)_{-N+N_f/2 } + N_f  \mbox{ real fermions}.
\end{align}
\end{subequations}
Both of these can be extended to the case of ``master dualities" with fermions and scalars on both sides \cite{Benini:2017aed,Jensen:2017bjo}; in any case, note that the assignments of the matrix size (which is now twice the rank) as well as the levels on the dual side exactly match the corresponding assignments in the unitary group. So going through the exercise of dualizing node by node we get exactly the same answers as in the unitary case, except for the fact that $SU$ and $U$ both get replaced with all $SO$ or all $Sp$ respectively. Let us focus here for simplicity on the case of unflavored quivers, the flavored case follows straightforwardly. We get the following two dualities in direct analogy with the unitary case:
\begin{equation}
\prod_{i=1}^n SO(N_i)_{-k_i}
\qquad\text{is dual to}\qquad
\left[\prod_{i=1}^{n-1} SO(K_i)_{N_i-N_{i+1}} \right]\times SO(K_n)_{N_n}
\end{equation}
and
\begin{equation}
\prod_{i=1}^n Sp(N_i)_{-k_i}
\qquad\text{is dual to}\qquad
\left[\prod_{i=1}^{n-1} Sp(K_i)_{N_i-N_{i+1}} \right]\times Sp(K_n)_{N_n}.
\end{equation}
Once again we can use these node-by node duality chains to argue for the special case that all $N_i=N$ and all ranks are $k_i=1$. In this case the original theories become quivers with equal rank and level nodes analogous to \ref{eq:su_quiver_ym}. In the dual theory we get all but the last node to be at level 0. Appealing to a confinement scenario as in the unitary case, we would argue that the last bifundamental gets replaced by a meson gauge invariant under the second to last node. This is a symmetric combination in the case of orthogonal gauge groups and an anti-symmetric combination in the symplectic case. Correspondingly we should conclude that
\begin{subequations}
\begin{align}
\label{eq:so_quiver_ym}
[SO(N)_{-1}]^n + \text{bifundamental scalars} \qquad\leftrightarrow\\
\label{eq:sod_quiver_ym}
 SO(n)_{N} + \text{symmetric rank-2 scalar}
\end{align}
\end{subequations}
and
\begin{subequations}
\begin{align}
\label{eq:sp_quiver_ym}
[Sp(N)_{-1}]^n + \text{bifundamental scalars}\qquad\leftrightarrow\\
\label{eq:spd_quiver_ym}
 Sp(n)_{N} + \text{antisymmetric rank-2 scalar}.
\end{align}
\end{subequations}
When we add flavors, we face same the complications as the unitary case. Fortunately these dualities can be extended into the flavor violated regime in exactly the same manner as the $SU$ theories. The main difference is the Grassmannians become
\begin{equation}
\frac{SO(N_f)}{SO(k)\times SO(N_f-k)}\qquad \text{and} \qquad \frac{Sp(N_f)}{Sp(k)\times Sp(N_f-k)} .
\end{equation}
These changes introduces a whole slew of interesting physics which is discussed in great detail in \cite{Komargodski:2017keh}. For our purposes, we simply note that the story of flavored quivers goes through in the same manner as in the $SU$ case.
\subsection{Phase Matching}
The phase matching in the symplectic and orthogonal cases also directly mimics their unitary counterparts. Let us focus first on the orthogonal case. In the quiver we can once again express the phases by partitions of $n$, determining whether the bifundamental scalars get positive or negative mass squareds. The generic phase of the quiver is given by
\begin{equation} \label{eq:sophases} \prod_i SO(N)_{- n_i} \end{equation}
in direct analogy with \eqref{eq:suphases} of the $SU$ quiver.
The phases of the $U(N)_n$ gauge theory were parametrized by the expectation values of the adjoint scalar, a hermitian matrix that we were able to diagonalize with the unitary gauge transformation. This time we are having a scalar in a symmetric matrix, exactly the object that can be diagonalized by our orthogonal gauge transformations. So once again the phases of the dual theory can be parametrized by the eigenvalues of the matrix and enhanced unbroken gauge groups arise whenever eigenvalues coincide. The phases of the dual theory hence become
\begin{equation} \label{eq:sodphases} \prod_i SO(n_i)_{N}, \end{equation}
indeed a level/rank dual representation of \eqref{eq:sophases}. In the symplectic case the antisymmetric matrix plays exactly the role of the symmetric matrix in the orthogonal case. It can be brought into normal form by symplectic transformations, breaking the gauge group down into products of smaller symplectic groups.

\subsection{Holographic Construction}

Holographic QCD was generalized to the orthogonal and symplectic case in \cite{Imoto:2009bf}. An orientifold O6 plane is introduced into the D4/D8 system in order to project the original unitary group down to either an orthogonal or symplectic subgroup. Both O6 and D8 are localized on the circle; if both are present they need to be offset from each other by an angle $\pi/4$ so that the stack of D8 branes gets mirrored onto the antipodal stack of anti-D8 branes by the orientifold projection. The two options for the gauge group are distinguished by exactly what type of O6 we add, the two options are usually denoted O6$^-$ for orthogonal and O6$^+$ for symplectic groups, the superscript indicating the sign of the RR charge associated with the orientifold plane. The full brane content both in the flat embedding space picture as well as in the dual geometry is enumerated in Tables \ref{branetablea} and \ref{branetableb}. Included in Table \ref{branetableb} is also the D6 brane that acts as the domain wall.
\begin{table}[h]
\begin{center}
\subfloat[Brane realization of the gauge theory in 10d flat space. $X_4$ is a compact direction. ]{
\begin{tabular}{c|cccccccccc}
\label{branetablea}
&0&1&2&3&4&5&6&7&8&9 \\
\hline
D4& \bf{x}& \bf{x}&\bf{x}&\bf{x}&\bf{x}&o&o&o&o&o \\
D8& \bf{x}&\bf{x}&\bf{x}&\bf{x}&o&\bf{x}&\bf{x}&\bf{x}&\bf{x}&\bf{x} \\
O6& \bf{x}&\bf{x}&\bf{x}&\bf{x}&o&\bf{x}&\bf{x}&\bf{x}&o&o
\end{tabular}}
\quad \hspace{1cm} \quad
\subfloat[Embedding of the probe branes in Witten's black hole.]{
\begin{tabular}{c|ccccccc}
\label{branetableb}
&0&1&2&3&$r$&$\theta$&$S^4$ \\
\hline
D8 &  \bf{x}&\bf{x}&\bf{x}&\bf{x}& \bf{x}& o& $S^4$ \\
O6 & \bf{x}&\bf{x}&\bf{x}&\bf{x}&\bf{x}&o& $S^2$ \\
D6 & \bf{x}&\bf{x}&\bf{x} &o& o&o & $S^4$
\end{tabular}}
\caption{Brane realization of holographic QCD with symplectic or orthogonal gauge group. In a) $N$ color D4 branes intersect an orientifold O6 to project to a real gauge group. In b) the same branes are embedded in the dual $\mathbb{R}^{3,1} \times D_2 \times S^4$ geometry, together with D6 branes dual to the $\theta$-interfaces. Also indicated are flavor D8 branes that would be needed to add additional flavors. $t$, $x_{1,2,3}$ are the coordinates on $\mathbb{R}^{3,1}$, and $r$ and $\theta$ the coordinates on $D_2$.}
\end{center}
\end{table}

In order to understand the theory in the bulk, we simply need to determine the effect of the O6 plane on the gauge theory living on the domain wall D6s. As is apparent from Tables \ref{branetablea} and \ref{branetableb} the O6 planes have 4 relative ND directions with both the D4s of the flat space embedding and the D6s in the black hole geometry. What this means is that the worldvolume theory on both D4s and D6s experience the same type of projection: the gauge groups is either orthogonal on both or symplectic on both. This is exactly as we would expect from our node-by-node procedure.

Last but not least we need to determine what happens to the formerly adjoint scalar corresponding to motion of the stack of D6s in the $x_3$ direction. Let us start with the case of an orthogonal gauge group. For a scalar describing motion transverse to the orientifold one finds that branes have to move off in mirror pairs, with a $U(1)$ gauge theory on each pair that can enhance to $U(n_i)$ for $n_i$ coincident branes on one side. This is the breaking pattern we would expect from an antisymmetric rank 2 tensor. On the other hand, a scalar corresponding to motion inside the orientifold allows the stack of $n$ branes to completely separate into $n$ individual branes - since they are inside the orientifold they need no mirror partner. Each brane has an $SO(1)$ gauge group on its worldvolume that can get enhanced to $SO(n_i)$ for coincident eigenvalues. Since the motion of the D6s into the $x_3$ direction is inside the O6, this corresponds to a symmetric rank-2 tensor. Reassuringly this is exactly what we are supposed to find according to our duality conjecture in \eqref{eq:sod_quiver_ym}. For symplectic gauge groups, the role of symmetric and antisymmetric rank-2 tensors is reversed. So the bulk physics yields indeed \eqref{eq:spd_quiver_ym} in that case.

In order to fully realize our dualities we would need to find \eqref{eq:so_quiver_ym} and \eqref{eq:sp_quiver_ym} to be the corresponding field theories living on the boundary. For the case of symplectic gauge groups this appears indeed to be the obvious guess for the theory on the domain walls of a confining $Sp(N)$ gauge theory generalizing the discussion of \cite{Gaiotto:2017tne} in the unitary case. In the orthogonal case this is certainly incorrect for the case of $SO(N)$ gauge groups, but it appears reasonable if the gauge group is instead $Spin(N)$. Since the stringy realizations of these gauge groups always involve heavy spinors, it is entirely reasonable to assume that the relevant gauge theories living on the branes were actually $Spin(N)$ groups all along. Unfortunately, these global issues are often not considered carefully in the orientifold literature. Somewhat confusingly, the node-by-node dualization above seems to be literally working with $SO$ groups, not $Spin$ groups. So while a coherent and self-consistent duality story seems to emerge on the symplectic side, a complete understanding of the orthogonal case would require solid control of these global issues.

\section{Discussion and Future Directions}
\label{sec:dfd}

In this work we extended the master duality presented in \cite{Jensen:2017bjo, Benini:2017aed} to the flavor violated regime where $N_f>k$.  It is important to keep in mind the general philosophy used in deriving the phase diagrams of Figs. \ref{fig: master_fv_n>ns} and \ref{fig: master_fv_n=ns}: the dynamics in the quantum phase is hidden by strong dynamics in the $SU$ theory. It is only when we pass to the $U$ theory that we can see interesting structure emerge, similar to approached used in \cite{Komargodski:2017keh} as well as in \cite{Gomis:2017ixy,Cordova:2017vab,Choi:2018tuh,Cordova:2018qvg}. Like these prior works, our dualities pass all the required consistency checks short of explicit large $N$ calculations. And although it is difficult to prove the existence of these new quantum phases exactly, 3d bosonization gives good evidence that they indeed exist.

Obvious generalizations of this approach include the flavor extension of the master duality to other regimes. Since one can effectively interchange the $SU$ and $U
$ sides of the duality by promoting the background global symmetry associated to $\tilde{A}_1$ to be dynamical, it should be straightforward to extend our work here to the $N_s>N$ and $N_f \leq k$ regime. The double flavor violated case, where $(N_f, N_s)>(k, N)$ would also be interesting to consider.\footnote{Recently, the authors of \cite{Armoni:2019lgb} analyzed the phase diagram of QCD$_3$ in the large $N$ limit under minimal assumption and found there should be sequence of first-order phase transitions coming from a multitude of metastable vacua. This phase structure is realized in the dual description with the help of potential term beyond quartic scalar term. It would also be interesting to see how these large $N$ features modify the analysis of the main text in more detail.}

Using the flavor violated master duality, we constructed a dual for a flavored quiver with bifundamental scalars on the links and fundamental fermions coupled to each node via node-by-node dualization. Such a theory describes interfaces in $3+1$ dimensional QCD as discussed in Ref. \cite{Gaiotto:2017tne}. The resulting dual theories were qualitatively supported by a holographic construction with D6 branes in the Sakai-Sugimoto model. We then extended our work to orthogonal and symplectic gauge groups, both in the node-by-node dualization and in holography. More interesting directions would be exploring other gauge theories with different matter contents. It is widely believed that infrared phase of three-dimensional theories describes effective worldvolume theory of walls/interfaces in corresponding four-dimensional theories \cite{Gomis:2017ixy,Acharya:2001dz,Komargodski:2017keh,Gaiotto:2017tne,Aitken:2018cvh,Argurio:2018uup,Choi:2018ohn,Bashmakov:2018ghn,Karasik:2019bxn}. It would be nice to understand what kind of quiver gauge theories with its phase diagram and stringy constructions would emerge when analyzing phase transitions of multiple walls/interfaces in various cases similar to what we've done in the case of QCD$_4$.

There are some limitations to our approach, however. First, the analysis in Appendix A of \cite{Gaiotto:2017tne} seems to indicate a periodicity of the Grassmannian as a function of the number of nodes, $n$. That is, the most general Grassmannian is not $\mathcal{M}(N_f,n)$ but instead $\mathcal{M}(N_f,n \mod N_f)$. Our quiver theory seems to indicate that this is not the case -- if the number of nodes $n$ is greater than $N_f$, the full Higgsed regime gives $SU(N)_{-n+\frac{N_f}{2}}$ with $N_f$ $\psi$. This is flavor bounded and so does not exhibit a quantum regime. The case examined in \cite{Gaiotto:2017tne} is valid for $n \le N_f$. The analysis just seems to point to this periodicity, but an explicitly demonstration of it has not yet been completed. Perhaps there is some subtle strong coupling dynamics in the $3+1$ dimensional theory that causes the Grassmannian to disappear. Or maybe our quiver theory is only valid for $n\leq N_f$ and there is some extension of our work that can bring us into the $n> N_f$ regime. We hope to report on this interesting question in future work.

We have also found that the quivers corresponding to $N_F=1$ and $N_F>1$ require the use of two very different versions of the master duality -- the double-flavor saturated and flavor violated cases, respectively. In Ref. \cite{Gaiotto:2017tne}, it was found these two cases also separated themselves quite distinctly due to the lack of an enhanced symmetry when the $3+1$-dimensional fermion mass disappeared in the $N_f=1$ case. It is not obvious to the authors if these two facts are somehow related.


The holographic side of things is ripe with interesting puzzles. For instance, it is still unknown what the exact form of the orthogonal duality is we derived in this manner. As briefly commented on in the end of Sec. \ref{sec:orth and symp}, the duality derived from the orientifold projection onto the orthogonal subgroup seems to be insensitive to the global properties of the gauge group. That is, we know that there are multiple different types of orthogonal gauge groups in 2+1 d that differ by the  required background terms and the gauging of various $\mathbb{Z}_2$ global symmetries \cite{Cordova:2017vab}. Can these global issues be understood in string theory? Moreover, how does one even see a quantum phase in the holographic construction of $SU(N)$ gauge theories with fundamental fermions in holography? We leave these questions for future work.

\section*{Acknowledgments}

We would like to thank Zohar Komargodski for useful discussions.
The work of KA, AB, and AK was supported, in part, by the U.S.~Department of Energy under Grant No.~DE-SC0011637.  Any opinions, findings, and conclusions or recommendations expressed in this material are those of the
authors and do not necessarily reflect the views of the funding agencies.

\begin{appendix}

\section{Background Terms and Flavor Violation}
\label{app:backgrounds}

In this appendix we work out the Lagrangians of the flavor extended master duality coupled to background fields, which provide important consistency checks for the phase diagram of the extended master duality proposed in section \ref{sec:fvm}. We mostly analyze the case of $N>N_s$ (Fig. \ref{fig: master_fv_n>ns}) and briefly comment for the case of $N=N_s$ (Fig. \ref{fig: master_fv_n=ns}). Finally we discuss gauging of flavor symmetry and description of flavor-violated quivers which are used extensively in the section \ref{sec:quivers}.

\subsection{$SU$ side}
First, we discuss the $SU$ side of the duality. The Lagrangian corresponding to the $SU$ side of flavor-violated master duality \eqref{flavor_violated_mast} is given by
\begin{align} \label{eq:app su lagrangian}
\mathcal{L}_{SU} & =\left|D_{b^{\prime}+B+\tilde{A}_1+\tilde{A}_2}\phi\right|^{2}+i\bar{\psi}\slashed{D}_{b^{\prime}+C+\tilde{A}_{1}}\psi+\mathcal{L}_{\text{int}}-i\left[\frac{N_{f}-k}{4\pi}\text{Tr}_{N}\left(b^{\prime}db^{\prime}-i\frac{2}{3}b^{\prime3}\right)\right]\nonumber \\
 & -i\left[\frac{N}{4\pi}\text{Tr}_{N_{f}}\left(CdC-i\frac{2}{3}C^{3}\right)+\frac{N(N_{f}-k)}{4\pi}\tilde{A}_{1}d\tilde{A}_{1}+2NN_{f}\text{CS}_{\text{grav}}\right]
\end{align}
which we denote using
\begin{equation}
SU\left(N\right)_{-k+N_{f}/2}\times\left[SU\left(N_{f}\right)_{N/2}\times SU\left(N_{s}\right)_{0}\times U\left(1\right)_{N(N_{f}-k)/2}\times U\left(1\right)_{0}\right]+N_f\psi+N_s\phi
\end{equation}
where again the terms in the $\left[\cdots\right]$ are global symmetries. In the mass deformed regime where $\left|m_{\psi}\right|\gg m_{*}$, phases are straightforwardly obtained from $SU$ side as follows:\footnote{We are suppressing gravitational Chern-Simons terms for brevity. They are straightforward to restore using level-rank duality.}
\begin{subequations}
\begin{align}
\text{(I)}:  & \qquad SU(N)_{-k+N_{f}}\!\times\!\left[SU\left(N_{f}\right)_{N}\!\times \!SU\left(N_{s}\right)_{0}\! \times \!J_I \right]\\
\text{(II)}:  & \qquad SU(N-N_{s})_{-k+N_{f}}\!\times\!\left[SU\left(N_{f}\right)_{N} \!\times \!SU\left(N_{s}\right)_{-k+N_{f}}\!\times\!  J_{II} \right]\\
\text{(III)}:  & \qquad SU(N-N_s)_{-k}\!\times\!\left[SU\left(N_{f}\right)_{0}\!\times \!SU\left(N_{s}\right)_{-k} \!\times\! J_{III}\right]\\[3pt]
\text{(IV)}:  & \qquad SU(N)_{-k}\!\times\!\left[SU\left(N_{f}\right)_{0}\!\times \!SU\left(N_{s}\right)_{0}\!\times \!J_{IV}\right].
\end{align}
\end{subequations}
where
\begin{equation}
\label{eq : U(1) definition appendix}
J_i \equiv J_i^{ab} \frac{1}{4\pi} \tilde{A}_a d \tilde{A}_b 
\end{equation}
is the Chern-Simons term for the Abelian background gauge fields in the $i$th phase in for $a,b = 1, 2$. For the asymptotic mass phases, the $J_i^{ab}$ are
\begin{subequations} \label{eq:appendix u1 first}
\begin{align}
 & J_{\text{I}}^{ab}=\begin{pmatrix} N(N_f-k) & 0 \\ 0 & 0 \end{pmatrix}
\\ & J_{\text{II}}^{ab}=\frac{N(N_f-k)}{N-N_s }\begin{pmatrix} N & N_s\\ N_s & N_s \end{pmatrix}
\\ & J_{\text{III}}^{ab}=\frac{-Nk}{N-N_s }\begin{pmatrix} N & N_s\\ N_s & N_s \end{pmatrix}
\\ & J_{\text{IV}}^{ab}= \begin{pmatrix} -Nk  & 0 \\ 0 & 0 \end{pmatrix}.
\end{align}
\label{eq:abelian factors new basis}
\end{subequations}

For the region of small fermion mass $\left|m_{\psi}\right|<m_{*}$, corresponding to phases $\text{V}$ to $\text{VIII}$, we are in the quantum regime and flavor symmetry is expected to be spontaneously broken by the non-perturbative fermion condensate, 
\begin{equation}
U(N_f) \rightarrow U(N_{f}-k)\times U(k).
\end{equation}
We now move on to the $U$ side of master duality, which describes quantum regime of $SU$ side as well as the semiclassical regimes

\subsection{$U$ side}

The Lagrangians for the $U$ side of the flavor-violated master duality can be fixed by demanding that they yield the same TFTs and background Chern-Simons terms as the $SU$ case for phases I to IV. As mentioned in the main text, this is achieved via two different scalar theories, one of which is identical to the flavor-bounded case. Explicitly, the Lagrangians corresponding to \eqref{flavor_violated_mast} are given by
\begin{subequations}
\begin{align} \label{eq:app u lagrangian}
\mathcal{L}_{U}^{m_{\psi}<0} & =\left|D_{c+C}\Phi_{1}\right|^{2}+i\bar{\Psi}_{1}\slashed{D}_{c+B+\tilde{A}_2}\Psi_{1}+\mathcal{L}_{\text{int}}^{\prime\left(1\right)}\nonumber \\
 & -i\left[\frac{N}{4\pi}\text{Tr}_{k}\left(cdc-i\frac{2}{3}c^3\right)-\frac{N}{2\pi}\text{Tr}_k(c)d\tilde{A}_1+ 2Nk\text{CS}_{\text{grav}}\right]
\\ \label{eq:app u lagrangian2}
\mathcal{L}_{U}^{m_{\psi}>0} & =\left|D_{c+C}\Phi_2\right|^2+i\bar{\Psi}_2\slashed{D}_{c+B-\tilde{A}_2}\Psi_2+\mathcal{L}_{\text{int}}^{\prime(2)}\nonumber \\
& -i\left[\frac{-N+N_s}{4\pi}\text{Tr}_{N_F-k}\left(cdc-i\frac{2}{3}c^3\right)+ 2\left(Nk -N_s(N_f-k)\right)\text{CS}_{\text{grav}}\right]\nonumber \\
& -i\left[\frac{N}{4\pi}\text{Tr}_{N_{f}}\left(CdC-i\frac{2}{3}C^3\right)+ \frac{N_f-k}{4\pi}\text{Tr}_{N_s}\left(BdB-i\frac{2}{3}B^3\right)\right] \nonumber \\
& -i\left[-\frac{N}{2\pi}\text{Tr}_{N_f-k}\left(c\right)d\tilde{A}_1- \frac{N_s}{2\pi}\text{Tr}_{N_f-k}(c)d\tilde{A}_2+\frac{N_s(N_f-k)}{4\pi}\tilde{A}_2 d\tilde{A}_2\right].
\end{align}
\end{subequations}
In general, dynamical gauge groups for these two theories are distinct, so the number of degrees of freedom in the matter fields is different. The reason the $m_\psi>0$ Lagrangian looks significantly more complicated is largely a result of our convention for $\eta$-invariant terms (see footnote 2). Schematically, these theories can be denoted by 
\begin{subequations}
\begin{align}
m_\psi<0:\qquad & U\left(k\right)_{N-N_{s}/2}\times\left[SU\left(N_{f}\right)_{0}\times SU\left(N_{s}\right)_{-k/2}\times U\left(1\right)_{NN_f/2}\times U\left(1\right)_{-NN_s/2}\right]\nonumber\\&+N_s \Psi_1+N_f\Phi_1 
\\
m_\psi>0:\qquad & U\left(N_{f}-k\right)_{-N+N_{s}/2}\times\left[SU\left(N_{f}\right)_{N}\times SU\left(N_{s}\right)_{\left(-k+N_{f}\right)/2}\times U\left(1\right)_{0}\times U\left(1\right)_{NN_s/2}\right]
\nonumber\\&+N_s \Psi_2+N_f\Phi_2.
\end{align}
\end{subequations}
Reassuringly, the procedure laid out in \cite{Jensen:2017bjo} gives background terms for the asymptotic region which match those from the $SU$ side. The next complication comes in determining the $U(1)$ levels in the quantum phase. On the $U$ side, the monopole current couples to $\tilde{A}_1$ which couples to the Kahler-form $w$
\begin{equation}
\mathcal{L}_U\supset  \frac{N}{2\pi} w d\tilde A_1.
\end{equation}
We can then integrate out $w$ in a procedure analogous to the semiclassical regions discussed above. 

As mentioned in the main text, moving into the quantum regions we find an additional $U(1)$ background symmetry which was a subgroup of the original $SU(N_f)$ global symmetry. Being careful with said breaking pattern results in the TFTs described in \eqref{eq:u phase n>ns}. 
\subsection{$N=N_s$}

We remark that all the above analysis of $N>N_s$ case is directly extended to $N=N_s$ with two important differences.

First, in the phases II and III there is a spontaneous breaking of the diagonal $ \tilde A_1 + \tilde A_2$ background field. This is straightforward to see on the $SU$ side where the dynamical gauge symmetry is completely Higgs when $N=N_s$. To see this on the $U$ side, one can move to a dual photon description to make the shift symmetry explicit. The breaking signals the naive divergence of background terms proportional to $\frac{N}{4\pi} (\tilde A_1 + \tilde A_2)d \tilde (\tilde A_1 + \tilde A_2 )$ in \eqref{eq:appendix u1 first} for $N=N_s$.

Second, as it's clear from the figure \ref{fig: master_fv_n=ns}, phases VII and VIII have finite regions in contrast to the $N>N_s$ case. Thus we have extra critical line II-III with $U(1)_0 +NN_f \text{ }\tilde \psi_s $, and it is reassuringly consistent with all the background terms calculated in phases II and III.

\subsection{Quiver description}

Since the flavor-violated $SU$ theory is master dual to two different $U$ theories which describe certain patches of the $SU$ phase diagram respectively, it will become useful to adopt a notation similar
to (\ref{eq:dual1 flq}) to denote the appropriate theory located
in the common region $m_{\psi}=0$ and $m_{\phi}^{2}=0$ of phase space, corresponding
to a Grassmannian manifold. One can choose either of the $U$ theories
to describe phases V and VI, which contain the $SU$ theory. Since
the two scalar theories are uniquely related to one another, later
we will adopt a method for converting from one scalar theory to the
other. As such, one should still be able to start with the $U$ theory
and determine all mass deformations. Choosing the $m_{\psi}<0$ theory,
we have a duality between the two theories
\begin{multline}
SU\left(N\right)_{-k+N_f/2}\times\left[SU\left(N_{f}\right)_{N_f/2}\times SU\left(N_{s}\right)_{0}\right] \qquad \leftrightarrow \\  U\left(k\right)_{N-N_s/2}\times\left[SU\left(N_{f}\right)_{0}\times SU\left(N_{s}\right)_{-k/2}\right].
\end{multline}

As with the flavorless case, it will be useful to have a version of
this theory where the $SU$ side has its fermion flavor symmetry and
a $U\left(1\right)$ global symmetry combined into a single $U\left(N_{f}\right)$
symmetry. This is achieved by grouping the $C$ and $\tilde{A}_{1}$
fields together(we do not include the additional $\tilde{B}$ symmetry to make contact with \ref{eq:dual1 flq} in addition to include $N=N_s$). The above duality becomes
\begin{multline}
SU\left(N\right)_{-k+N_{f}/2}\times\left[U\left(N_{f}\right)_{N/2}\times SU\left(N_{s}\right)_{0}\right] \qquad \leftrightarrow \\  U\left(k\right)_{N-N_{s}/2}\times\left[U\left(N_{f}\right)_{0}\times SU\left(N_{s}\right)_{-k/2}\right].\label{eq:master ext neg flq}
\end{multline}
Immediately note that this duality is identical to that of (\ref{eq:dual3 flq}).

Alternatively, if we had chosen the $m_{\psi}>0$ theory on the $U$
side, we would have arrive at something a little different, namely,
\begin{multline}
SU\left(N\right)_{-k+N_{f}/2}\times\left[U\left(N_{f}\right)_{N/2}\times SU\left(N_{s}\right)_{0}\right] \qquad \leftrightarrow \\U\left(N_{f}-k\right)_{-N+N_{s}/2}\times\left[U\left(N_{f}\right)_{N}\times SU\left(N_{s}\right)_{\left(-k+N_{f}\right)/2}\right].\label{eq:master ext pos flq}
\end{multline}

\section{Quivers and $\text{Spin}_{c}$}
\label{sec:Quivers and spinc flq}

In \cite{Aitken:2018joi} it was discussed how the master duality
is consistent with being put on a $\text{spin}_{c}$ manifold. Unfortunately,
we weren't particularly careful with ordinary and $\text{spin}_{c}$
connections in \cite{Aitken:2018cvh}, so we should clarify the
consistency here.

In \cite{Aitken:2018cvh} a modified version of the master duality is used and various shifts were performed. Specifically, a shift on the Abelian portion of $c$ was performed, $\tilde{c}\to\tilde{c}+\tilde{A}_{1}$ and the common $\tilde{A}_{1}$ Chern-Simons term was canceled. Being more careful with connections, this is equivalent to redefining a new $\text{spin}_{c}$ gauge field which we will call $\tilde{a}=\tilde{c}-\tilde{A}_{1}$ and ordinary gauge field $\tilde{B}=\tilde{A}_{1}+\tilde{A}_{2}$, so that the $U$ side is now
\begin{align}
\mathcal{L}_{U} & =\left|D_{c^{\prime}+\tilde{a}+\tilde{A}_{1}+C}\Phi\right|^{2}+i\bar{\Psi}\slashed{D}_{c^{\prime}+\tilde{a}+B+\tilde{B}}\Psi+\mathcal{L}_{\text{int}}^{\prime}\nonumber \\
 & -i\left[\frac{N}{4\pi}\text{Tr}_{k}\left(c^{\prime}dc^{\prime}-i\frac{2}{3}c^{\prime3}\right)+\frac{Nk}{4\pi}\tilde{a}d\tilde{a}+2Nk\text{CS}_{\text{grav}}\right].
\end{align}
We also shifted the $\tilde{A}_{2}$ fields into the $\psi$ and $\Phi$
matter by taking $\tilde{A}_{1}\to\tilde{A}_{1}-\tilde{A}_{2}$. Instead,
define a new ordinary connection $\tilde{G}=\tilde{A}_{1}+\tilde{A}_{2}$(corresponds to the $\tilde B$ in the $U$ side above)
so the $SU$ side of the duality now reads
\begin{align}
\mathcal{L}_{SU} & =\left|D_{b^{\prime}+B+\tilde{G}}\phi\right|^{2}+i\bar{\psi}\slashed{D}_{b^{\prime}+C+\tilde{A}_{1}}\psi+\mathcal{L}_{\text{int}}-i\left[\frac{N_{f}-k}{4\pi}\text{Tr}_{N}\left(b^{\prime}db^{\prime}-i\frac{2}{3}b^{\prime3}\right)\right]\nonumber \\
 & -i\left[\frac{N}{4\pi}\text{Tr}_{N_{f}}\left(CdC-i\frac{2}{3}C^{3}\right)+NN_{f}\left(\frac{1}{4\pi}\tilde{A}_{1}d\tilde{A}_{1}\right)+2NN_{f}\text{CS}_{\text{grav}}\right].
\end{align}
This is the same as we have in our paper with the replacements, $\tilde{G}\to\tilde{A}_{1}$,
$\tilde{A}_{1}\to\tilde{A}_{1}-\tilde{A}_{2}$, $c^{\prime}+\tilde{a}\to c$,
$\tilde{B}\to\tilde{A}_{1}$. We also promote the field $\tilde{A}_{2}$
which introduces a new background symmetry which we call $\tilde{B}_{2}$.
In our new notation this is equivalent to gauging $\tilde{A}_{1}\to\tilde{a}_{1}$,
and we will called the new ordinary connection to which it couples
$\tilde{B}_{1}$.

The part we need to be most careful is in the match of the two theories
as we move across the quiver construction. This is spelled out in the most detail in Appendix A.3 of \cite{Aitken:2018cvh}
become. There we make certain identifications
based on fermion interactions which are now equivalent to
\begin{equation}
i\bar{\Psi} { D \! \! \! \! \slash }_{c^{\prime}+\tilde{a}+B+\tilde{B}}\Psi\qquad\leftrightarrow\qquad i\bar{\psi} { D \! \! \! \! \slash }_{b^{\prime}+C+\tilde{A}_{1}}\psi.
\end{equation}
The field level identification which are in Table 3 of \cite{Aitken:2018cvh}
become
\begin{subequations}
\begin{align}
c^{\prime}+\tilde{a}\qquad & \leftrightarrow\qquad C+\tilde{a}_{1}\\
B\qquad & \leftrightarrow\qquad b^{\prime}\\
\tilde{B}\qquad & \leftrightarrow\qquad\tilde{B}_{1}.
\end{align}
\end{subequations}
Since we are matching ordinary connections to ordinary connections
and $\text{spin}_{c}$ connections to $\text{spin}_{c}$ connections,
everything is consistent and it does appear these quivers are consistent
with being put on $\text{spin}_{c}$ manifolds.

\end{appendix}

\bibliographystyle{JHEP}
\bibliography{flavorquivers}
\end{document}